\PassOptionsToPackage{hyphens}{url}
\documentclass[aps,prx,twocolumn,longbibliography]{revtex4-1}

\usepackage{amsmath}
\usepackage{amsfonts}
\usepackage{mathtools}
\usepackage{braket}
\usepackage{bm}
\usepackage{bbm}
\usepackage{array}
\usepackage{graphicx}
\usepackage{amssymb}
\usepackage[dvipsnames]{xcolor}
\usepackage{tikz-cd}

\usepackage{hyperref}
\usepackage{bbold}

\hypersetup{
	colorlinks=true,
	linkcolor=blue,
	citecolor=blue,
	filecolor=magenta,
	urlcolor=cyan,
	breaklinks=true
}
\DeclareMathOperator{\Tr}{Tr}

\DeclareMathOperator{\im}{im}

\newcommand{\papertitle}{Interacting Symmetry-Protected Topological Phases Out of Equilibrium}
\newcommand{\tcm}{T.C.M. Group, Cavendish Laboratory, University of Cambridge, JJ Thomson Avenue, Cambridge, CB3 0HE, U.K.}

\DeclareSymbolFont{sfletters}{OML}{cmbrm}{m}{it}

\newcommand{\iu}{{\rm i}}

\DeclareMathSymbol{\matrrho}{\mathord}{sfletters}{"1A}
\newcommand{\diff}{\mathrm{d}}

\DeclareFontFamily{OMX}{MnSymbolE}{}
\DeclareFontShape{OMX}{MnSymbolE}{m}{n}{
	<-6>  MnSymbolE5
	<6-7>  MnSymbolE6
	<7-8>  MnSymbolE7
	<8-9>  MnSymbolE8
	<9-10> MnSymbolE9
	<10-12> MnSymbolE10
	<12->   MnSymbolE12}{}
\DeclareSymbolFont{mnlargesymbols}{OMX}{MnSymbolE}{m}{n}
\SetSymbolFont{mnlargesymbols}{bold}{OMX}{MnSymbolE}{b}{n}
\DeclareMathDelimiter{\llangle}{\mathopen}{mnlargesymbols}{'164}{mnlargesymbols}{'164}
\DeclareMathDelimiter{\rrangle}{\mathclose}{mnlargesymbols}{'171}{mnlargesymbols}{'171}

\makeatletter
\newcommand*{\itemequation}[2][]{%
	\item
	\begingroup
	\refstepcounter{equation}%
	\ifx\\#1\\%
	\else
	\label{#1}%
	\fi
	#2
	\begin{sloppypar}
		\hfill \@eqnnum %
	\end{sloppypar}
	\endgroup
}
\makeatother

\begin{document}
	
	\title{\papertitle}
	\author{Max McGinley}
	\affiliation{\tcm}
	\author{Nigel R. Cooper}
	\affiliation{\tcm}
	
	\date{\today}
	
	\begin{abstract}
		
		
		The topological features of quantum many-body wave functions are known to have profound consequences for the physics of ground-states and their low-energy excitations. We describe how topology influences the dynamics of many-body systems when driven far from equilibrium. Our results are succinctly captured by a non-equilibrium topological classification that can be used to predict universal aspects of generic isolated quantum systems as they evolve unitarily in time. By analogy to the classifications used to describe systems in equilibrium, we consider two short-ranged entangled wavefunctions to be topologically equivalent if they can be interconverted via finite-time unitary evolution governed by a symmetry-respecting Hamiltonian. We demonstrate that this definition captures the salient features of these systems in a broad range of non-equilibrium scenarios. As well as providing conceptual insights into the constraints imposed by topology on many-body dynamics, we discuss the practical implications of our findings. In particular, we show that the characteristic zero-frequency spectroscopic peaks associated with topologically protected edge modes will be broadened by external noise only when the system is trivial in the non-equilibrium classification.
%
%
%
	\end{abstract}
	
	\maketitle

	\section{Introduction}
	
	Since the theoretical prediction \cite{Kane2005,Kane2005a,Fu2007} and subsequent experimental observation \cite{Konig2007,Hsieh2008} of topological insulators in two and three spatial dimensions, there has been enormous interest in understanding the role of topology in band insulators and superconductors. A milestone in this field was the classification of all topological phases that can be described by non-interacting fermions with non-spatial symmetries \cite{Kitaev2009,Ryu2010}. One of the attractive features of these phases is their robustness to weak interactions, which are inevitably present in any experimental setting.
	
	However, when one enters the regime of strong many-body interactions, two phenomena must be accounted for. On the one hand, some free-fermion phases become trivial in the many-body context, leading to a reduction in their classification \cite{Fidkowski2011,Tang2012,Wang2014}. On the other, strong interactions open up the possibility of realizing new topological phases which have no non-interacting analogue. A long-known example 
	occurs in integer-spin chains, which host a topologically non-trivial phase of matter known as the Haldane phase \cite{Haldane1983,Haldane1983a,Haldane1985}.
	
	In understanding exactly how these phases differ from their trivial counterparts, a useful unifying framework has emerged which naturally incorporates both non-interacting and strongly interacting systems. In this framework, the aforementioned examples are all \textit{symmetry-protected topological phases} (SPTs): each possesses a ground state which cannot be connected to a trivial state by adiabatically varying the Hamiltonian without closing the gap or explicitly breaking the relevant symmetries \cite{Chen2010}. This is a natural construction for characterizing systems at zero temperature; however, it is reasonable to ask whether a different construction may be useful in other scenarios.
	
	\begin{figure}[b]
		\includegraphics[scale=1]{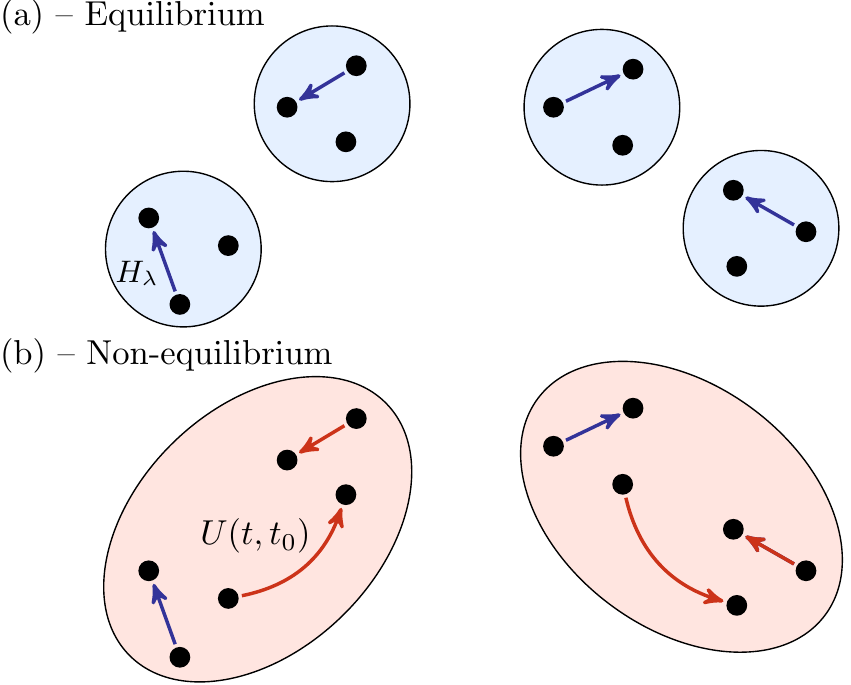}
		\caption{Illustration of equilibrium vs non-equilibrium topological classification. Black dots represent various short-ranged entangled wavefunctions that respect a certain symmetry group $G$. Panel (a) -- In equilibrium, wavefunctions are classified into sets (blue circles) according to whether each can be adiabatically connected through a family of symmetry-respecting Hamiltonians $H_\lambda$ parametrised by $\lambda \in [0,1]$ (blue arrows). Panel (b) -- The non-equilibrium classification partitions wavefunctions into sets (red ellipses) according whether each can be connected through some finite-time unitary evolution $U(t,t_0) = \mathcal{T} \exp[-\iu\int_{t_0}^t \diff t' H(t')]$ ($\mathcal{T} = \text{time-ordering}$) governed by a Hamiltonian $H(t')$ that respects the symmetries in $G$ (red arrows). 
		}
		\label{figClass}
	\end{figure}
	
	The dynamics of systems far from equilibrium is one scenario that cannot be captured by a ground state. Such far-from-equilibrium dynamics is of particular current interest in light of recent developments of a wide variety of experimental platforms in which the coherent quantum dynamics of many-particle systems can be studied. Recently, theoretical progress has been made in generalising notions of topology to this setting. This includes studying periodically driven Hamiltonians, whose Floquet eigenstates can be topologically characterised \cite{Kitagawa2010,Rudner2013,Else2016, vonKeyserlingk2016, Potter2016,Roy2017}, as well as identifying fingerprints of static topological phases in quench dynamics \cite{Wang2017,Tarnowski2017,Zhang2018,Sun2018,Gong2018,Shuangyuan2019}. In addition to these works, which describe features of the system's trajectory over time, the instantaneous topological properties of wavefunctions undergoing time evolution have also been studied \cite{DAlessio2015,Caio2015}.
	 In this approach, the time dependence of non-interacting bulk indices can be understood, allowing one to systematically characterise non-interacting free fermion systems far from equilibrium \cite{McGinley2018,McGinley2019}. However, the techniques applied in these previous studies are specific to non-interacting systems, and do not generalize to strongly interacting SPT phases.
	
	
	In this paper, we provide a framework for classifying the topology of many-body wavefunctions using a construction that applies naturally to non-equilibrium scenarios, and that can be generalized beyond free fermions. In place of the equilibrium approach, in which one differentiates between phases according to whether they can be adiabatically deformed between each other via a series of gapped symmetric Hamiltonians, here we consider wavefunctions to be equivalent if they can be deformed into each other via \textit{finite-time unitary evolution governed by a (possibly time-dependent) Hamiltonian that respects the relevant symmetries}. The sub-extensive evolution time ensures that this constitutes a relationship between short-ranged entangled wavefunctions, and thus plays the role of the bulk gap in equilibrium. As illustrated in Fig.~\ref{figClass}, one can construct equivalence classes under this relation, which constitute the non-equilibrium topological classification.
	
	\begin{table}
		\begin{ruledtabular}
			\begin{tabular}{lcc}
				Symmetry group & Equilibrium & Non-eq. \\
				\colrule Bosonic: & & \\
				$\mathbb{Z}_2^T$ & $\mathbb{Z}_2$ & $0$ \\
				$\mathbb{Z}_2^T \times \text{trn}$ & $\mathbb{Z}_2$ & $0$ \\
				$U(1) \times \mathbb{Z}_2^T$ & $\mathbb{Z}_2 \times \mathbb{Z}_2$ & $0$ \\
				$U(1) \times \mathbb{Z}_2^T \times \text{trn}$ & $\mathbb{Z}_2 \times \mathbb{Z}_2$ & 0 \\
				$U(1) \rtimes \mathbb{Z}_2^T$ & $\mathbb{Z}_2 \times \mathbb{Z}_2$ & $0$ \\
				$U(1) \rtimes \mathbb{Z}_2^T \times \text{trn}$ & $\mathbb{Z} \times \mathbb{Z}_2$ & $\mathbb{Z}$ \\
				$\mathbb{Z}_n \times \mathbb{Z}_2^T$ & $\mathbb{Z}_2 \times \mathbb{Z}_{(2,n)}$ & $0$ \\
				$\mathbb{Z}_n \rtimes \mathbb{Z}_2^T$ & $\mathbb{Z}_2 \times \mathbb{Z}_{(2,n)}$ & $0$ \\
				$\mathbb{Z}_n \times \mathbb{Z}_m \times \mathbb{Z}_2^T$ & $\mathbb{Z}_2 \times \mathbb{Z}_{(2,n)} \times \mathbb{Z}_{(2,m)} \times \mathbb{Z}_{(2,n,m)}$ & $\mathbb{Z}_{(2,n,m)}$ \\
				$SO(3) \times \mathbb{Z}_2^T$ & $\mathbb{Z}_2 \times \mathbb{Z}_2$ & $\mathbb{Z}_2$ \\
				\colrule Fermionic: & & \\
				$\mathbb{Z}_2^F \times \mathbb{Z}_2^T$ (BDI) & $\mathbb{Z}_8$ & $\mathbb{Z}_2$ \\
				$\mathbb{Z}_4^{T,F}$ (DIII) & $\mathbb{Z}_2$ & $0$ \\
				$U(1) \times \mathbb{Z}_2^{T}$ (AIII) & $\mathbb{Z}_4$ & 0
			\end{tabular}
		\end{ruledtabular}
		\caption{Non-equilibrium classifcation of 1D fermionic and bosonic interacting SPT phases protected by onsite symmetry group $G = G_T \times \mathbb{Z}_2^T$ or $G = G_T \rtimes \mathbb{Z}_2^T$, given by the data \eqref{enumNoneq}. The group $G_T$ is realised unitarily and $\mathbb{Z}_2^T = {e,T}$, where $T$ represents time-reversal, and is realised antiunitarily. $\mathbb{Z}_4^{T,F}$ is the cyclic group $\{1,T,P_f,TP_f\}$, representing time-reversal symmetry and fermion parity $P^f$ for half-integer spin fermions. Here `trn' denotes translational invariance, and $(n,m)$ is the greatest common divisor of integers $n$ and $m$. The non-equilibrium classification for symmetry groups only featuring unitary elements are identical to the equilibrium classification, which can be found in Ref.~\cite{Chen2013}. Important bosonic symmetry classes with are bosonic topological superconductors ($\mathbb{Z}_2^T$), bosonic topological insulators ($U(1) \rtimes \mathbb{Z}_2^T$), and $S^z$-conserving spin chains without a magnetic field ($U(1) \times \mathbb{Z}_2^T$). We also include some fermionic systems which are interacting analogues of the non-interacting symmetry classes within the ten-fold way; the Cartan labels for the corresponding Altland-Zirnbauer class are also given in brackets.}
		\label{tab1D}
	\end{table}
	
	The `deformations' which one is permitted to make to identify wavefunctions in this classification are more general than those permitted by the more familiar equilibrium classification. Adiabatic evolution through a series of gapped symmetric Hamiltonians ensures that the wavefunction respects all symmetries at all times. However, the same is not necessarily true for unitary evolution under a symmetric Hamiltonian. It may be possible to construct a unitary evolution between two symmetric wavefunctions where the wavefunction at intermediate times does not respect the symmetry of the governing Hamiltonian. This phenomenon -- which we have called dynamically induced symmetry breaking  \cite{McGinley2018} -- leads to a reduced non-equilibrium classification compared to that in equilibrium.
	
	When this framework is applied to free-fermion systems, previously established results concerning the dynamics of topological bulk invariants can be understood. Specifically, the bulk indices of the time-evolved state (as studied in Ref.~\cite{McGinley2019})
	provide labels for those non-equilibrium topological classes that have a non-interacting representation. The new construction presented here has the advantage that it naturally generalizes to interacting systems, where the non-interacting bulk indices do not apply. In this case, to explicitly compute the classification, one must adopt the theoretical technology used to describe short-ranged entangled many-body wavefunctions, namely matrix-product states \cite{Fannes1992,PerezGarcia2007} (MPS) and their generalization to higher dimensions \cite{Verstraete2004,Verstraete2006}.
	
	\bgroup
	\def\arraystretch{1.25}
	\begin{table*}
		\begin{ruledtabular}
			\begin{tabular}{lcccc}
				Symmetry group & \multicolumn{4}{c}{Spatial dimension $d$} \\
				& $0$ & $1$ & $2$ & $3$ \\
				\colrule $\mathbb{Z}_2^T$ & $0$ & $\mathbb{Z}_2$ & $0$ & $\mathbb{Z}_2$ \\
				& $0$ & $0$ & $0$ & $0$ \\
				\colrule $U(1) \times \mathbb{Z}_2^T$ & $0$ & $\mathbb{Z}_2^{\times2}$ & $0$ & $\mathbb{Z}_2^{\times3}$ \\
				& $0$ & $0$ & $0$ & $0$ \\
				\colrule $U(1) \rtimes \mathbb{Z}_2^T$ & $\mathbb{Z}$ & $\mathbb{Z}_2$ & $\mathbb{Z}_2$ & $\mathbb{Z}_2^{\times2}$ \\
				& $\mathbb{Z}$ & $0$ & $0$ & $0$ \\
				\colrule $\mathbb{Z}_n \times \mathbb{Z}_2^T$ & $\mathbb{Z}_{(2,n)}$ & $\mathbb{Z}_2 \times \mathbb{Z}_{(2,n)}$ & $\mathbb{Z}_{(2,n)}^{\times2}$ & $\mathbb{Z}_2 \times \mathbb{Z}_{(2,n)}^{\times2}$ \\
				& $\mathbb{Z}_{(2,n)}$ & $0$ & $\mathbb{Z}_{(2,n)}$ & $0$ \\
				\colrule $\mathbb{Z}_n \rtimes \mathbb{Z}_2^T$ & $\mathbb{Z}_{n}$ & $\mathbb{Z}_2 \times \mathbb{Z}_{(2,n)}$ & $\mathbb{Z}_{(2,n)}^{\times2}$ & $\mathbb{Z}_2 \times \mathbb{Z}_{(2,n)}^{\times2}$ \\
				& $\mathbb{Z}_{n}$ & $0$ & $\mathbb{Z}_{(2,n)}$ & $0$ \\
				\colrule $\mathbb{Z}_n \times \mathbb{Z}_m \times \mathbb{Z}_2^T$ & $\mathbb{Z}_{(2,n)} \times \mathbb{Z}_{(2,m)}$ & $\mathbb{Z}_2 \times \mathbb{Z}_{(2,n)} \times \mathbb{Z}_{(2,m)} \times \mathbb{Z}_{(2,n,m)}$ & $\mathbb{Z}_{(2,n)}^{\times2} \times \mathbb{Z}_{(2,m)}^{\times2} \times \mathbb{Z}_{(2,n,m)}^{\times2}$ & $\mathbb{Z}_2 \times \mathbb{Z}_{(2,n)}^{\times2} \times \mathbb{Z}_{(2,m)}^{\times2} \times \mathbb{Z}_{(2,n,m)}^{\times4}$ \\
				& $\mathbb{Z}_{(2,n)} \times \mathbb{Z}_{(2,m)}$ & $\mathbb{Z}_{(2,n,m)}$ & $\mathbb{Z}_{(2,n)} \times \mathbb{Z}_{(2,m)} \times \mathbb{Z}_{(2,n,m)}$ & $\mathbb{Z}_{(2,n,m)}^{\times2}$ \\
				\colrule $SO(3) \times \mathbb{Z}_2^T$ & $0$ & $\mathbb{Z}_2^{\times2}$ & $\mathbb{Z}_2$ & $\mathbb{Z}_2^{\times3}$ \\
				& $0$ & $\mathbb{Z}_2$ & $0$ & $0$ \\
			\end{tabular}
		\end{ruledtabular}
		\caption{Non-equilibrium classification of bosonic interacting symmetry-protected topological orders in physical spatial dimensions, as captured by Eq.~\eqref{eqClass}. Systems are protected by various onsite symmetry groups $G = G_T \times \mathbb{Z}_2^T$ or $G = G_T \rtimes \mathbb{Z}_2^T$, where $G_T$ is realised unitarily and $\mathbb{Z}_2^T = {e,T}$, where $T$ represents time-reversal, and is realised antiunitarily. Here $(n,m)$ is the greatest common divisor of integers $n$ and $m$. The non-equilibrium classification for symmetry groups without the time-reversal part are identical to the equilibrium classification, which can be found in Ref.~\cite{Chen2013}. For each symmetry group, we provide two rows, the first being the equilibrium classification taken from Ref.~\cite{Chen2013}, and the second being the non-equilibrium classification discussed in Section \ref{secHigher} derived in Appendix \ref{secSS}. Systems with translational invariance also possess weak indices, which out of equilibrium are still given by products of the non-equilibrium classification in lower spatial dimensions.}
		\label{tabAllDim}
	\end{table*}
	\egroup
	
	Through studying all possible symmetry-respecting MPS wavefunctions a classification scheme SPT phases in equilibrium has been presented. The equilibrium classification in $d$ spatial dimensions is given by the $(d+1)$th cohomology group $\mathcal{H}^{(d+1)}[G,U_T(1)]$ \cite{Chen2013} (although this analysis misses some exotic `beyond-cohomology' phases in higher dimensions \cite{Vishwanath2013}).
	By utilizing these cohomological techniques, we identify the differences in the structure of MPS and related states before and after some generic time evolution, and distinguish structures that can or cannot change during the evolution process. This gives us the means to compute the non-equilibrium classification explicitly, which we present for systems with various physically relevant symmetries in Tables \ref{tab1D} and \ref{tabAllDim}.

	Thanks to the naturalness and generality of its construction, the non-equilibrium classification has many physical consequences which are reflected in the dynamics of systems which exhibit SPT order. In particular, quantities which are relevant for topological phases in equilibrium exhibit universal behaviour in non-equilibrium scenarios. For example, after some generic time evolution starting in an SPT phase, one can identify whether the entanglement spectrum will remain gapless or become gapped \cite{Gong2019}, and whether string order parameters remain non-zero \cite{Calvanese2016,Hagymasi2019,Hagymasi2019a}. These information-theoretic quantities naturally witness the changes of topology in the many-body wavefunction which occur under unitary evolution. We also establish a link between the classification and more directly observable quantities, namely spectroscopy measurements, which are used to identify gapless modes at the boundaries of SPT phases in equilibrium. Under very general conditions, we show that the presence of low-frequency classical noise will only broaden these characteristic spectral peaks when the system belongs to a trivial class in the non-equilibrium classification. These results are highlighted in the context of a recent experiment realising an SPT phase in a chain of Rydberg atoms \cite{Leseleuc2018}.

	

	
	In section \ref{secProj}, we review how projective symmetry representations can be used to classify one dimensional (1D) SPT phases in equilibrium. We then apply these methods to construct the 1D non-equilibrium classification in Section \ref{secNoneq1D}. Section \ref{secHigher} uses the same guiding principles to calculate the non-equilibrium classification in higher dimensions, making use of the group cohomology methods introduced in Ref.~\cite{Chen2013}. We discuss the consequences of our results for directly observable quantities in Section \ref{secExpt}, before concluding in Section \ref{secConc}.

	\section{Projective representations and 1D SPT phases\label{secProj}}
	
	We begin by reviewing how interacting symmetry-protected topological phases of bosons can be understood and classified in 1D systems through studying projective representations of the symmetry group, which was first described in Refs.~\cite{Pollmann2010,Chen2011,Schuch2011,Chen2011a,Pollmann2012}. We consider the ground state wavefunctions of local 1D gapped bosonic systems which do not spontaneously break the symmetry of the Hamiltonian. As is well known, such states are well approximated by matrix-product states (MPS) due to their area-law entanglement \cite{Hastings2007,Schuch2008}. The MPS ansatz for the ground state takes the form \cite{Fannes1992,PerezGarcia2007}
	\begin{align}
	\ket{\Psi} = \sum_{i_1, i_2, \ldots, i_L} \Tr \left[A^{[1]}_{i_1}A^{[2]}_{i_2} \cdots A^{[N]}_{i_N}\right]\ket{i_1, i_2, \cdots, i_N}.
	\label{eqMPS}
	\end{align}
	Here, $\ket{i_1, i_2, \cdots, i_N}$ is a product state labelled by the on-site quantum numbers $\{i_k = 1,\ldots,s\}$ for local dimension $s$, and each $A^{[k]}_{i_k}$ is a $D\times D$ matrix that parametrizes the wavefunction. Although the bond dimension $D$ is arbitrary, a wavefunction with area law entanglement can be efficiently captured using some fixed finite choice of $D$.

	Different quantum phases are understood to be captured by ground state wavefunctions that cannot be continuously connected between one another via local unitary operations \cite{Chen2010}. States that can be connected to a trivial reference state in this way are termed short-ranged entangled (SRE).  In 1D, if a gapped Hamiltonian has a unique ground state, then it is necessarily SRE; thus all non-degenerate ground states belong to the same phase. The states we consider in this paper have no spontaneously broken symmetries, and will all be SRE. This implies certain (generalized) injectivity conditions on the relevant wavefunction ansatz, which we implicitly assume from hereon (See Refs.~\cite{Chen2011,Williamson2016} for details). 
	
	However, if the Hamiltonian respects a symmetry that is not spontaneously broken, then wavefunctions can be considered as inequivalent if they cannot be continuously connected via local \textit{symmetry-respecting} unitary operations \footnote{We distinguish a unitary operator $\hat{U}$ which respects the symmetries from a unitary operator $e^{-i\hat{H}t}$ generated by a symmetry respecting Hamiltonian}. The physical relevance of this construction to the zero-temperature properties of many-body systems becomes apparent when one interprets this unitary operation as an adiabatic evolution through a series of gapped symmetric Hamiltonians. A set of wavefunctions which are all mutually equivalent in this sense constitutes an SPT phase. 
	
	If one insists that the wavefunction respects a symmetry with symmetry group $G$, then this imposes restrictions on the $A_{i_k}^{[k]}$. If the symmetry is realised by an on-site unitary representation on the Hilbert space $u(g)\, \otimes\, u(g)\, \otimes\, \cdots\, \otimes\, u(g)$, where each unitary $u(g)$ is a $s\times s$ matrix acting on one site only, then the condition for $\ket{\Psi}$ to respect the symmetry is \cite{Chen2011a}
	\begin{align}
	\sum_{j=1}^s u(g)_{ij} A_j = \alpha(g)R^{-1}(g)A_i R(g)
	\label{eqMPSUnit}
	\end{align}
	where $\alpha(g)$ is a scalar, $R(g)$ is a $D \times D$ matrix, and we have assumed translational invariance, dropping the superscript labels on the $A$. Importantly, to be consistent with the group structure of $G$, the matrices $R(g)$ need only respect the multiplication rule of $G$ up to a phase factor \cite{Pollmann2010}: $R(g_1g_2) = \omega(g_1, g_2)R(g_1)R(g_2)$, where $\omega(g_1, g_2)$ is a modulus-1 complex number known as the factor system. This implies that the $R(g)$ form a \textit{projective representation} of the symmetry group $G$. It was shown \cite{Chen2011} that wavefunctions belong to the same SPT phase \textit{if and only if} their projective representations are equivalent, in the sense that they can be related through multiplying one by a linear representation. [This equivalence relation is captured by the second cohomology group of $G$: $\omega \in \mathcal{H}^{(2)}(G,\mathbb{C})$, see Sec.~\ref{secHigher}]. Note also that $\alpha(g)$, which must form a 1D linear representation of $G$, is a good quantum number only when translational invariance is imposed, whilst the structure of $R(g)$ persists even when translational invariance is broken \cite{Chen2011}. Therefore, when the ground state respects a on-site unitary symmetry with group $G$, SPT phases are characterized by
	\begin{enumerate}
		\item The equivalence class of the projective representation of $G$, $\omega(g_1, g_2)\in H^2(G,\mathbb{C})$
		\itemequation[enumSPT]{ (Translational invariance only) The 1D linear representation $\alpha(g) \in \mathcal{G}$.}
	\end{enumerate}
	Here, $\mathcal{G}$ is the group of 1D linear representations of $G$.
	
	The analysis when $G$ features antiunitary (time-reversal) elements follows similarly. In this case, $G$ will factorize as $G = G_T \times \mathbb{Z}_2^T$ or $G = G_T \rtimes \mathbb{Z}_2^T$, where the group $\mathbb{Z}_2^T = \{1, T\}$ contains the time-reversal symmetry (TRS) generator, and the choice depends on whether time-reversal commutes with the other group elements or not \footnote{Most generally, $G$ is a group extension of its unitary subgroup $G_T$ by an antiunitary group $\mathbb{Z}_2^T$: $1 \rightarrow G_T \rightarrow  G \rightarrow \mathbb{Z}_2^T \rightarrow 1$ which might not split as a direct or semi-direct product \cite{Moore2014}. A notable exception is half-integer-spin fermionic systems with a time-reversal symmetry satisfying $T^2 = P_f$ ($P_f$ is the fermion parity operator), which has a $\mathbb{Z}_4^T$ structure $\{I,T,P_f,TP_f\}$.}.
	Time-reversal $T$ acts on the Hilbert space as $V(T) = v\, \otimes\, v\, \otimes\, \cdots\, \otimes\, v\, K$, where $K$ is the complex conjugation operator and $v$ is an on-site unitary. For the bosonic spin systems considered in the majority of this paper, one can assume that $vv^* = 1$ without loss of generality. (In a system for which $vv^* = -1$, the unit cell can be doubled, after which one has $vv^* = 1$ \cite{Chen2011}.)
	Just as for unitary symmetries, the MPS can be classified according to the way in which it transforms under the action of the symmetries. The unitary subgroup $G_T$ generates the same data as described above. In addition to this, the MPS must also transform consistently under the action of the antiunitary element $T$. This implies that
	\begin{align}
	\sum_{j=1}^s v_{ij}A^*_j = M^{-1}A_i M
	\label{eqMPSTRS}
	\end{align}
	for some $D \times D$ matrix $M$. To be consistent with the $\mathbb{Z}_2^T$ group product $T^2 = 1$, we must have $MM^* = \beta(T)\mathbb{1}$, where $\beta(T) = \pm 1$ \cite{Chen2011a}. The two choices of $\beta(T)$ capture the different projective representations of $\mathbb{Z}_2^T$.
	
	The objects so far $\omega(g_1,g_2)$, $\alpha(g)$, $\beta(T)$ quantify how the group product is represented for $G_T$ and $\mathbb{Z}_2^T$ separately; however one also needs to understand how group products between unitary and antiunitary elements are realised. By applying the symmetries \eqref{eqMPSUnit} and \eqref{eqMPSTRS} in different orders, the authors of Ref.~\cite{Chen2011a} demonstrated that, when $G = G_T \times \mathbb{Z}_2^T$ (i.e.~$T g T = g$), the projective representations must satisfy a projective commutation relation
	\begin{align}
	M^{-1} R(g) M = \gamma(g)R(g)^*
	\end{align}
	where $\gamma(g)$ is a phase factor that forms a linear 1D representation of $G_T$. The representation $\gamma(g)$ is only uniquely determined up to multiplication by some other 1D representation which is the square of another 1D representation [which can be absorbed into $R(g)$]. Different elements of $\mathcal{G}/\mathcal{G}_2$ therefore represent distinct SPT phases, where $\mathcal{G}$ is the group of 1D representations of $G_T$, and $\mathcal{G}_2$ is the group of those representations which are squares of other representations. When $G = G_T \rtimes \mathbb{Z}_2^T$, one replaces $R(g)$ with $R(g^{-1})$ on the right-hand side to account for the fact that elements of $G_T$ do not commute with time-reversal.
	
	On the one hand, this new relation, which follows from the presence of an additional antiunitary symmetry $T$, serves to extend the possible topological phases compared to the unitary case, since different representations $\gamma(g)$ correspond to different phases. On the other hand, combining this relation with \eqref{eqMPSUnit}, one finds that $\omega(g_1, g_2)$ and $\alpha(g)$ are restricted to square to unity, which implies that some of the phases that existed under the symmetry $G_T$ are not compatible with the presence of TRS (specifically, those with $\omega^2 \neq 1$ or $\alpha^2 \neq 1$). To summarise, different topological phases under a on-site symmetry group $G_T \times \mathbb{Z}_2^T$ featuring a time-reversal part are
	\begin{enumerate}
		\item The equivalence class of the projective representation of $G$, $\omega(g_1, g_2)$, {which must satisfy} $\omega^2 = 1$
		\item (Translational invariance only) The 1D linear representation $\alpha(g)$, {which must satisfy} $\alpha^2 = 1$.
		\item {The projective representation of} $\mathbb{Z}_2^T$, $\beta(T) = \pm 1$
		\itemequation[enumTRS]{ {The projective commutation relation between symmetry transformations in $G_0$ and $\mathbb{Z}_2^T$, $\gamma(g) \in \mathcal{G}/\mathcal{G}_2$}}
	\end{enumerate}
	Note that while the above refers to bosonic models, fermionic models can also be captured using the same arguments through a Jordan-Wigner transformation \cite{Chen2011a}. In that context, one must also include phases which result from a spontaneous breaking of the $\mathbb{Z}_2^P$ fermion parity symmetry, which results in boundary Majorana modes. 
	
	Understanding the way that TRS both enhances and restricts the topological classification compared to the case where $G = G_T$ will be central to our understanding of these phases when out of equilibrium.
	
	\section{Non-equilibrium SPT classification in 1D \label{secNoneq1D}}
	
	The methods described in the previous section allow one to explicitly compute the equilibrium topological classification for 1D bosonic systems \cite{Chen2010,Chen2011a}. The data regarding linear and projective representations of the symmetry group [Eqs.~\eqref{enumSPT}, \eqref{enumTRS}] fully determine whether one ground state can be deformed into another via adiabatic evolution along a path of symmetry-respecting Hamiltonians [Fig.~\ref{figClass}(a)]. In this section, we describe how these techniques can be further applied to compute the non-equilibrium classification, which specifies whether one symmetry-respecting short-ranged entangled wavefunction can be deformed into another via finite-time unitary evolution governed by a symmetry-respecting Hamiltonian [Fig.~\ref{figClass}(b)]. 
	
	The key difference between the equilibrium and non-equilibrium deformation procedures is the possibility that unitary (non-adiabatic) evolution between symmetry-respecting SRE wavefunctions $\ket{\Psi_1}$ and $\ket{\Psi_2}$ may proceed via states which at intermediate times do not respect all the symmetries of the Hamiltonian \cite{McGinley2018}. In particular, those symmetries that are realised antiunitarily (e.g.~time reversal symmetry) are generically broken under unitary evolution; we refer to this as \textit{dynamically induced symmetry breaking}. One can understand this since the factor of $\rm i$ in the time-evolution operator is not invariant under an antiunitary operator: $T e^{-\iu\hat{H} t} T^{-1} = e^{+\iu\hat{H} t}$. (Note that our non-equilibrium construction allows the Hamiltonian to vary in time itself, so long as it remains symmetry-respecting; in that case a similar equation can be derived for the time evolution operator). 
	
	We specify that the unitary evolution occurs over a finite time; however this condition should be made precise. We require that the time evolution can be accurately captured by some finite-depth unitary circuit. For a generic system with a local Hamiltonian, this implies that the evolution time is less than $L_\text{sys}/v_{\text{L.R.}}$, where $L_\text{sys}$ is the system size and $v_{\text{L.R.}}$ is a Lieb-Robinson velocity. The locality properties of an adiabatic deformation and of a finite-time evolution are therefore equivalent, since the former can also be captured by a finite-depth unitary circuit \cite{Chen2010}. Importantly, in both cases the wavefunction is always amenable to a matrix product state description (albeit with a bond dimension that typically grows exponentially in time \cite{Prosen2007}). As such, the methods described in the previous section can be readily applied to the non-equilibrium construction.
	
	One must therefore distinguish between a unitary circuit that is itself symmetry-respecting (adiabatic deformation) from a unitary circuit generated by a symmetry-respecting Hamiltonian (finite-time unitary evolution). The latter is strictly more general than the former, which implies that the non-equilibrium classification will be a subgroup of the equilibrium classification.

	With the above in mind, let us first consider a system which respects a group of on-site unitary symmetries $G = G_T$. If two $G_T$-symmetric wavefunctions $\ket{\Psi_1}$ and $\ket{\Psi_2}$ are in the same non-equilibrium topological class, then by definition one can specify a unitary time evolution $\ket{\Psi(t)}$ satisfying $\ket{\Psi(0)} = \ket{\Psi_1}$, $\ket{\Psi(1)} = \ket{\Psi_2}$. The initial state $\ket{\Psi_1}$ will be topologically characterized by the data described in Eqs.~\eqref{enumSPT}. Because all symmetries are realised unitarily and the time evolution is over a finite time, the wavefunctions at intermediate times respect all the same symmetries and remain SRE, and thus can be characterized in the same way. Since the topological data are discrete and well-defined for times $0 \leq t \leq 1$, they cannot change under such a unitary evolution. Thus if $\ket{\Psi_1}$ and $\ket{\Psi_2}$ belong to different equilibrium topological phases, then such a unitary path is impossible. We conclude that the non-equilibrium topological classification for systems with only unitary symmetries is the same as the familiar equilibrium classification.
	
	Now consider the case where the system also respects a time-reversal symmetry $G = G_T \times \mathbb{Z}_2^T$, so that the initial state is characterized by the data in Eq.~\eqref{enumTRS}. (The $G_T \rtimes \mathbb{Z}_2^T$ case proceeds in the same way.) Because of dynamically induced symmetry breaking, 
	the wavefunction at intermediate times will only respect the symmetries in $G_T$. This means that for $0 < t < 1$, some of the topological data used to characterize the states at $t = 0, 1$ may not be well-defined. In particular, we expect that $\beta(T)$ and $\gamma(g)$, each of which relate to the matrix $M$ in \eqref{eqMPSTRS}, become meaningless. Therefore, even if $\ket{\Psi_1}$ and $\ket{\Psi_2}$ have different $\beta(T), \gamma(g)$ [but the same $\omega(g_1, g_2)$ and $\alpha(g)$], they can still be connected via finite-time unitary evolution under a symmetry-respecting Hamiltonian. If they differ with respect to $\omega, \alpha$, which are both still well-defined along the trajectory, then they cannot be connected in this way, and belong to different non-equilibrium classes.
	
	The objects that can still distinguish wavefunctions in the non-equilibrium classification are simply those that characterise the equilibrium topology of systems that only respect the symmetry subgroup $G_T$. Note, however, that the non-equilibrium classification is not simply given by the equilibrium classification of this reduced symmetry group, since the states $\ket{\Psi_1}$, $\ket{\Psi_2}$ must belong to phases that are compatible with the full symmetry group $G_T \times \mathbb{Z}_2^T$. From Eq.~\eqref{enumTRS}, we see that TRS requires the representations to satisfy $\omega^2 = 1$, $\alpha^2 = 1$, which comprise only a subgroup of the equilibrium classification for $G_T$-symmetric systems.
	
	In summary, the non-equilibrium classification comprises of all possible combinations of the following data:
	\begin{enumerate}
		\item The equivalence class of the projective representation of $G$, $\omega(g_1, g_2)$, which must satisfy $\omega^2 = 1$
		\itemequation[enumNoneq]{ (Translational invariance only) The 1D linear representation $\alpha(g)$ which must satisfy $\alpha^2 = 1$.}
	\end{enumerate}
	
	The non-equilibrium classification we describe in this paper is formally equivalent to the construction we presented in Refs.~\cite{McGinley2018, McGinley2019} when applied to free fermions. In those works, rather than describing equivalences between two symmetry-respecting SRE wavefunctions, we demonstrated how to capture the topological properties of a wavefunction after some generic quench protocol, starting from the ground state of a symmetry-respecting Hamiltonian. Again in that context, it is the dynamical breaking of antiunitary symmetries which reduces the classification compared to in equilibrium. The previous approach clearly shows how our results are of direct applicability to quench protocols, however our new construction is closely analogous to the equilibrium classification, and has an intrinsic relevance to non-equilibrium protocols beyond quenches. Directly observable physical consequences of the non-equilibrium classification therefore emerge in these scenarios -- we discuss some of these consequences in Section \ref{secExpt}.

	Having described how the non-equilibrium classification can be constructed for interacting SPT phases in 1D, we tabulate some examples for physically relevant symmetry groups in Table \ref{tab1D}. In many cases, the non-equilibrium classification is trivial, since the equilibrium classification of $G_T$ (without the TRS part) is itself trivial. More subtle cases include the time-reversal invariant spin chains with full rotation invariance $SO(3) \times \mathbb{Z}_2^T$. In this case, the equilibrium classification is $\mathbb{Z}_2 \times \mathbb{Z}_2$; the first group factor accounting for $\beta(T) = \pm 1$ and the second accounting for the two projective representations of $SO(3)$ [which are integer-spin and half-integer-spin linear representations of its double cover $SU(2)$]. Only the latter object is well-defined once TRS is dynamically broken, and so the non-equilibrium classification reduces from $\mathbb{Z}_2 \times \mathbb{Z}_2$ to $\mathbb{Z}_2$.
	
	We can also make connection with our previous results on non-equilibrium classifications of free-fermion systems \cite{McGinley2018,McGinley2019} using the Jordan-Wigner transform approach \cite{Chen2011a}. In that context, the fermion systems belong to one of 10 Altland-Zirnbauer symmetry classes \cite{Altland1997}, which can then be re-interpreted as symmetry groups of the auxiliary spin system (although in some cases one needs to specify whether the free Hamiltonian represents a superconductor or an insulator \cite{Morimoto2015}). Spinless superconductors with time-reversal symmetry are in class BDI in the free case and possess a $\mathbb{Z}_2^F \times \mathbb{Z}_2^T$ symmetry group in the many-body language ($\mathbb{Z}_2^F$ represents fermion parity, whilst $\mathbb{Z}_2^T$ is time-reversal symmetry with $T^2 = +1$). Our finding in Ref.~\cite{McGinley2018} that the non-interacting time-evolved state is only characterised by the $\mathbb{Z}_2$ subgroup of the equilibrium classification $\mathbb{Z}$ is consistent with the interacting picture. When interactions are added, the equilibrium classification is reduced from $\mathbb{Z}$ to $\mathbb{Z}_8$, consisting of one trivial phase, 3 phases where $T, P_f$ are unbroken, but TRS is projectively realised, and 4 phases where the fermion parity symmetry is \textit{spontaneously} broken, and $T$ is realised linearly or projectively. After TRS is dynamcially broken, the first 3 non-trivial phases are all indistinguishable from the trivial phase, and the latter 4 phases remain non-trivial, but mutually indistinguishable, hence the $\mathbb{Z}_2$ non-equilibrium classification, the same as in the non-interacting case. We find similar agreement for spinful time-reversal symmetric superconductors with (AIII) and without (DIII) $U(1)$ spin-rotation invariance (because $T^2 = P_f$, we use $\mathbb{Z}_4^{T,F}$, the cyclic group $\{I,T,P_f,TP_f\}$).
	
	\section{Extending to higher dimensions \label{secHigher}}
	
	The arguments regarding projective representations of the symmetry groups are specific to 1D, since they rely on the matrix-product state ansatz \eqref{eqMPS}. To generalize to higher dimensional systems whose ground states possess area-law entanglement, one must consider more general tensor-network states, such as projected entangled-pair states (PEPS)  \cite{Verstraete2004,Verstraete2006}. As described in Ref.~\cite{Chen2013}, the natural algebraic structure for classifying bosonic SPT phases in this context is the cohomology group $\mathcal{H}^{1+d}[G,U_T(1)]$, where $d$ is the spatial dimension , and $U_T(1)$ is defined below. (Note that this construction excludes systems with mixed gauge-gravity anomalies and surface intrinsic topological order \cite{Vishwanath2013,Wang2013,Burnell2014,Wen2015}; we do not specifically analyse those systems here, but expect that the same principles can be used to analyse them.) A full introduction to group cohomology and its relevance to SPT phases can be found in Ref.~\cite{Chen2013}; however, we will briefly summarize its structure here.
	
	The principles underpinning how SPT phases are classified in higher dimensions are no different from 1D, but some technical aspects of the arguments are altered. It turns out to be more convenient not to explicitly separate TRS from the unitary symmetries as we did above, (which previously resulted in additional data $\beta$, $\gamma$). This can be achieved by specifying how the symmetry transformations $g \in G$ act on wavefunctions, such that the antiunitary nature of TRS is captured. To be specific, each element $g \in G$ acts on complex phases $a \in U(1)$ via the product $g \cdot a$ such that $g \cdot a = a^*$ if $g$ is antiunitary, and $g \cdot a = a$ otherwise. $U_T(1)$, which consists of the abelian group $U(1)$ combined with the operation $\cdot$, is referred to as a $G$-module.
	
	The key object used in Section \ref{secProj} was the factor system of the projective representation $\omega(g_1, g_2)$, which quantifies how the representation fails to respect the group product $R(g_1)R(g_2) = \omega(g_1, g_2)R(g_1g_2)$. If TRS is included within $G$, then the factor system $\omega : G^2 \rightarrow U_T(1)$ is formally a map from 2 group elements to the $G$-module $U_T(1)$ which is the only object needed to specify an SPT phase. The structure of these maps can be understood in the framework of group cohomology, which more generally concerns maps from $n$ group elements to an arbitrary $G$-module $M$. The space of such maps is denoted $\mathcal{C}^n[G,M] = \{\omega : G^n \rightarrow M\}$.
	
	Now define a `differential' operator $d_n : \mathcal{C}^{n}[G,M] \rightarrow \mathcal{C}^{n+1}[G,M]$ which returns a function of $(n+1)$ elements of $G$
	\begin{align}
	&(d_n\omega)(g_1, \ldots , g_{n+1}) = [g_1 \cdot \omega(g_2, \ldots, g_{n+1})] \nonumber\\
	&\times \omega^{(-1)^{n+1}}(g_1,\ldots,g_n)\prod_{i=1}^n \omega^{(-1)^i}(g_1,\ldots,g_{i-1}, g_i g_{i+1},\ldots g_n).
	\end{align}
	The precise form of the differential operator is not important; we need only know that $d_n$ is a homomorphism, and $d_{n+1} \circ d_n = 0$. This last identity justifies the nomenclature, since the same identity is satisfied by the exterior derivative in differential geometry.
	
	One considers the infinite family of groups $\{\mathcal{C}^{n}[G,M] : n \geq 0\}$ (where we understand $\mathcal{C}^{0}[G,M] = M$) along with the maps $\{d_n\}$ between them, which together constitute a \textit{cochain complex}
	\begin{align}
	\mathcal{C}^0[G,M] \xrightarrow{d_0} \mathcal{C}^1[G,M] \xrightarrow{d_1} \mathcal{C}^2[G,M] \xrightarrow{d_2} \mathcal{C}^3[G,M] \xrightarrow{d_3} \cdots
	\label{eqCochain}
	\end{align}
	An element of $\mathcal{C}^n[G,M]$ is referred to as an $n$-cochain.
	
	If an $n$-cochain $\omega_n$ can be written $\omega_n = d_{n-1}\omega_{n-1}$ for some $\omega_{n-1}$ (i.e.~$\omega_n \in \im d_{n-1}$), one says that it is exact, and if it satisfies $d_n \omega_n = 0$ (i.e.~$\omega_n \in \ker d_n$), then one says that it is closed. From $d_{n+1} \circ d_n = 0$, all exact $n$-cochains are closed.
	However, not all closed cochains are exact. The $n$th cohomology group of this complex quantifies this asymmetry; it is defined as the quotient group
	\begin{align}
	\mathcal{H}^n[G,M] \coloneqq \ker (d_{n}) / \im (d_{n-1}),
	\label{eqCohom}
	\end{align}
	or in words, `the equivalence classes of cochains that are closed, but cannot be interconverted through multiplying by an exact cochain'. If, for a particular $G$-module $M$, all closed chains are exact, then every cohomology group is trivial, and the sequence \eqref{eqCochain} is a long exact sequence.
	
	Returning to the 1D case, the natural quantities to deal with are elements of $\mathcal{C}^2[G,U_T(1)]$. Note, however, that not all such functions from $G^2$ to $U_T(1)$ are valid projective representations. One can think of the condition that $\omega \in \ker (d_2)$ as a statement that $\omega$ is the factor system of a valid projective representation. Two projective representations $\omega, \omega'$ are considered to be equivalent if they are related by a 1D linear representation through $\omega(g_1, g_2) = \omega'(g_1, g_2)\beta(g_2)^{s(g_1)}\beta(g_1)/ \beta(g_1 g_2)$ for some $\beta \in \mathcal{C}^1[G,U_T(1)]$ \cite{Chen2010}, since such a change can be absorbed into $\alpha(g)$ in Eq.~\eqref{eqMPSUnit}. Here, $s(g_1) = \pm 1$ depending on whether $g_1$ is an antiunitary element. This equivalence of $\omega, \omega'$ can be written as $\omega = \omega' \times (d_1 \beta)$, which means that they belong to the same class in $\mathcal{H}^2[G,U_T(1)]$, and represent the same SPT phase. Conversely, if $\omega$ and $\omega'$ are in different classes of $\mathcal{H}^2[G,U_T(1)]$, then they correspond to different SPT phases. For higher dimensions, different SPT phases are captured by elements of $\mathcal{H}^{1+d}[G,U_T(1)]$ \cite{Chen2013}.
	
	We now describe how the above construction of cohomology groups can be applied to our non-equilibrium classification. As in 1D (see Sec.~\ref{secNoneq1D}), we must consider a unitary evolution $\ket{\Psi(t)}$ between symmetric SRE wavefunctions $\ket{\Psi_1}$ at $t=0$, and $\ket{\Psi_2}$ at $t=1$. Dynamically induced symmetry breaking means that at intermediate times $0 < t < 1$, $\ket{\Psi(t)}$ will respect only a subgroup of the symmetries $G_T \leq G$, in which only unitary elements are kept. We must understand which topological data remain well-defined throughout the evolution, since only these data will restrict whether $\ket{\Psi_1}$ and $\ket{\Psi_1}$ can be unitarily connected.

	At time $t=0$, the object $\omega_0 \in \mathcal{C}^{n}[G,U_T(1)]$, which belongs to one of the equivalence classes of $\mathcal{H}^{n}[G,U_T(1)]$, characterizes how an initial PEPS state $\ket{\Psi_1}$ (or other wavefunction ansatz) transforms under the full symmetry group $G$ in dimension $d = n-1$. However, $\ket{\Psi(t>0)}$ can only be understood through its behaviour under symmetry transformations within the subgroup $G_T$. We should therefore take the function $\omega_0 : G^n \rightarrow U_T(1)$ and restrict it to the domain $G_T^n$, yielding $\omega_T \coloneqq \omega_0|_{G_T}$. This object is sufficient to characterize the topology of $\ket{\Psi(t)}$, but does not do so uniquely. For the same reasons as in equilibrium, we must identify how $\omega_T$ fits into the cohomology group corresponding to the reduced set of symmetries.
	
	Note that $\omega_T$ is an element of $\mathcal{C}^{n}[G_T,U(1)]$ [we drop the subscript $T$ on the module since by definition all elements of $G_T$ are unitary and have trivial action on $U_T(1)$]. The groups $\mathcal{C}^{n}[G_T,U(1)]$ form their own cochain complex, and for each $n$ we can define the restriction map $\text{res}_n : \mathcal{C}^{n}[G,U(1)] \rightarrow \mathcal{C}^{n}[G_T,U(1)]$, defined as above. Importantly, the restriction map is a homomorphism, and the following diagram is commutative \cite{Sharifi}
	\begin{widetext}
		\begin{equation}
		\begin{tikzcd}
		\mathcal{C}^0[G,U_T(1)] \ar[r,"d_0"] \ar[d,"\text{res}_0"] & \mathcal{C}^1[G,U_T(1)] \ar[d, "\text{res}_1"] \ar[r,"d_1"] & \mathcal{C}^2[G,U_T(1)] \ar[d, "\text{res}_2"] \ar[r,"d_2"] & \cdots \\
		\mathcal{C}^0[G_T,U(1)] \ar[r, "d_0^T"] & \mathcal{C}^1[G_T,U(1)] \ar[r,"d_1^T"] & \mathcal{C}^2[G_T,U(1)] \ar[r,"d_2^T"] & \cdots
		\end{tikzcd}
		\end{equation}
	\end{widetext}
	i.e.~restriction preserves which elements are exact and which are closed. Here, we use $d_n^T$ to denote the differential maps on the bottom cochain complex. It is well known in cohomology that this restriction from $G$ to any subgroup in turn induces a homomorphism on the cohomology groups, called the restriction functor \cite{Sharifi}
	\begin{align}
	\text{Res}_n : \mathcal{H}^n[G,U_T(1)] \rightarrow \mathcal{H}^n[G_T,U(1)].
	\label{eqResFunctor}
	\end{align}
	We use a capitalized $\text{Res}_n$ to denote the restriction functor on cohomology groups.
	
	To construct $\text{Res}_n$ explicitly, one can consider how the restriction map $\text{res}_n$ affects the components of Eq.~\eqref{eqCohom}. First, consider $\ker(d^{n})$, i.e.~the group of closed $n$-cochains within $\mathcal{C}^{n}[G,U_T(1)]$. Since the differentials and restrictions commute, any closed element of $\mathcal{C}^{n}[G,U_T(1)]$ will still be closed when restricted to an element of $\mathcal{C}^{n}[G_T,U(1)]$. Restriction therefore defines a map between the two kernels $\text{res}_{n} : \ker (d_{n}) \rightarrow \ker (d_{n}^T)$. Note, however, that not all closed elements of $\mathcal{C}^{n}[G_T,U(1)]$ can be written as the restriction of some closed element of $\mathcal{C}^{n}[G,U_T(1)]$.
	
	For $n=2$ ($d = 1$), this is the statement that a valid projective representation of $G$ becomes a valid projective representation of $G_T$ after restriction, but that not all projective $G_T$-representations can be extended to projective $G$-representations. In the context of higher-dimensional SPT phases (for $n > 2$), one can replace the notion of a valid projective $G$-representation with a valid action of the symmetry $G$ on the wavefunction ansatz of choice, as specified by some $\omega_{n} \in \mathcal{C}^{n}[G,U_T(1)]$. The previous paragraph simply states that if $\ket{\Psi_1}$ transforms in a consistent way under $G$, then it must also transform consistently under $G_T$, but that the converse is not necessarily true.
	
	Similarly, thanks to the commutative diagram above, $\text{res}_n$ maps elements of $\im(d_{n-1})$ to elements of $\im(d_{n-1}^T)$, i.e.~exact cochains remain exact after restriction. Again, not all exact cochains of the reduced group $G_T$ can be expressed as restrictions of exact cochains of $G$. Two cochains $\omega_1, \omega_2$ which are inequivalent as elements of $\mathcal{H}^{n}[G,U_T(1)]$ must satisfy $\omega_1 \neq \omega_2 \times (d_{n-1} \beta)$ for all $\beta \in \mathcal{C}^{n-1}[G,U_T(1)]$. However, after restriction to $G_T$ they may become equivalent as elements of $\mathcal{H}^{n}[G_T,U(1)]$, since one may have $\omega_1 = \omega_2 \times (d_{n-1}^T \beta_T)$ for $\beta_T \in \mathcal{C}^{n-1}[G_T,U_T(1)]$.
	
	
	The restriction functor $\text{Res}_n$ between the cohomology groups defined in \eqref{eqResFunctor} is constructed through the action of $\text{res}_n$ on $\ker(d_{n+1})$ modulo the transformations defined by $\im(d_n^T)$, in the sense defined above. It has an image
	\begin{align}
	\im\left(\text{Res}_n\right) = \im \left(\ker(d_{n+1}) \xrightarrow{\text{res}_n} \ker(d_{n+1}^T) \right) / \im (d_n^T).
	\label{eqClass}
	\end{align}
	The above object \eqref{eqClass} constitutes the non-equilibrium classification for bosonic systems in spatial dimension $d = n-1$. Each element of this group represents a collection of symmetry-respecting SRE wavefunctions which can be mutually connected via finite-time unitary evolution under a symmetry-respecting Hamiltonian.
	
	In Appendix \ref{secSS} we describe a method for computing the image of the restriction map by making use of the Hochschild-Serre spectral sequence \cite{Babakhanian1972}. The non-equilibrium classification for all pure bosonic SPT phases computed using this method is given in Table \ref{tabAllDim}. To provide some insight as to how the non-equilibrium classification manifests itself, as well as to verify our results, in Appendix \ref{secExamples} we provide two contrasting examples of quench protocols in 2D interacting SPT phases which can be solved exactly.

	\section{Physical consequences of the non-equilibrium classification \label{secExpt}}
	
	The familiar equilibrium classification of SPT phases, which is based on adiabatic deformations, is extremely powerful in predicting robust features of interacting systems which are in their ground state. For example, a topologically non-trivial ground state universally possess a gapless entanglement spectrum \cite{Li2008,Pollmann2010}, as well as long range string order \cite{Nijs1989,Kennedy1992,Haegeman2012}. Additionally, topologically protected boundary modes are expected to appear in SPT phases, which can be easily probed in experiment using spectroscopic techniques \cite{Leseleuc2018}.
	
	The non-equilibrium classification derived in the preceding sections is constructed in a similar way, but is based on deformations generated by unitary time evolution rather than adiabatic evolution. In an analogous way to equilibrium, the classification has directly observable physical consequences for interacting systems which feature various forms of external time dependence. We highlight some of these effects in this section.
	
	Our construction most obviously lends itself to quench protocols. In that context, one can look for universal features of the time-evolved wavefunction, as witnessed by those quantities that are used to identify topologically non-trivial ground states in equilibrium. One can, for instance, consider the entanglement spectrum of a wavefunction following some quench. We expect that the entanglement spectrum will generically be gapless only if the initial state is non-trivial under the non-equilibrium classification. This is because states which are trivial under the non-equilibrium classification by definition can be locally deformed to a product state without breaking any symmetries (since the antiunitary symmetries are already broken after the quench), and will therefore not exhibit topologically protected entanglement degeneracies. The same will be true for long range string order. Furthermore, the possibility of accessing many-body SPT invariants in experiment using randomized measurements \cite{Elben2019} would allow the loss of SPT order after a quench to be directly observed. These predictions are consistent with previous work on quenches in free-fermion systems \cite{McGinley2018,McGinley2019}, which is natural since our classification can be applied to such systems in a way consistent with those studies.
	
	The non-equilibrium classification also applies to protocols in which the external time-dependence is non-deterministic, e.g.~classical noise. In the context of topological quantum computation, noise-induced decoherence of qubits stored in Majorana bound states has previously been studied \cite{Schmidt2012,Konschelle2013,Hu2015}.
	It is natural to expect more generally that SPT phases in the presence of classical noise exhibit universal phenomenology related to our non-equilibrium construction. We will show that our results are indeed reflected in spectroscopy measurements of SPT systems when subjected to low-frequency noise -- specifically, we find that the zero-energy peaks associated with boundary degeneracies will be broadened if the system is trivial in the non-equilibrium classification, and remain sharp if the system is non-trivial. This behaviour is highlighted using an example model used to describe a recent Rydberg atom experiment \cite{Leseleuc2018}. We then discuss in a broader sense the contexts in which our classification scheme captures the key effects.
	
	\subsection{Effect of classical noise on edge mode spectroscopy \label{secNoise}}
	
	In one spatial dimension $d=1$, SPT systems in equilibrium with open boundaries possess ground state degeneracies that can be probed using spectroscopic techniques. A coarse model for an SPT system with a single boundary involves two many-body ground states $\ket{0}$ and $\ket{1}$, and a set of excited states $\{\ket{n}\}$ with energies above the bulk gap $\epsilon_n \geq E_\text{g}$. (A particular SPT phase may have an edge degeneracy of more than two, but the physics remains the same.) The two ground states differ only near the edge, and thus can be connected by a local operator; however such an operator must violate some of the protecting symmetries. The spectral function of such an operator, which is measured in spectroscopy, will include a zero-frequency component, signalling the degeneracy of $\ket{0}$ and $\ket{1}$.
	
	We will show that when the unperturbed Hamiltonian is subjected to low-frequency classical noise, the characteristic edge mode peak of the $d=1$ SPT system will broaden only if the original system is trivial in the non-equilibrium classification. Although an analysis for SPT systems with an arbitrary symmetry group $G$ is in principle possible, we find it more instructive to consider two example cases, protected by symmetry groups $\mathbb{Z}_2 \times \mathbb{Z}_2$, and $\mathbb{Z}_2 \times \mathbb{Z}_2^T$. The former has only unitary symmetries, and is therefore non-trivial out of equilibrium, whilst the latter features antiunitary elements and has a trivial classification out of equilibrium. Both groups are of relevance to the experiment in Ref.~\cite{Leseleuc2018}, which we discuss in Section \ref{secRyd}.
	
	 In each case, the Hilbert space (of dimension $N$) forms a representation of the relevant symmetry group. In a non-trivial phase, the ground state subspace forms an irreducible subrepresentation, which is the origin of its protection. We choose a basis in which the action of the symmetries on this degenerate space $\{\ket{0}, \ket{1}\}$ takes the form
	\begin{align}
	&\{\mathbbm{1}_2,\, \hat{\tau}^x,  \iu\hat{\tau}^y,\,  \hat{\tau}^z\}& (\mathbb{Z}_2 \times \mathbb{Z}_2)& \nonumber\\
	&\{\mathbbm{1}_2,\, \hat{\tau}^x \hat{\mathcal{K}},\, \iu \hat{\tau}^y \hat{\mathcal{K}},\,  \hat{\tau}^z\}& (\mathbb{Z}_2 \times \mathbb{Z}_2^T)&,
	\label{eqSymOps}
	\end{align}
	where $\hat{\tau}^{x,y,z}$ are Pauli matrices and $\hat{\mathcal{K}}$ is the complex conjugation operator. One can verify that each set of operators generates an irreducible representation, so that the only allowed Hamiltonian term in the subspace is proportional to $\mathbbm{1}_2$.
	
	
	Spectroscopy of the system involves measuring a connected two-time correlation function
	\begin{align}
	\Gamma_B(t) \coloneqq \left\langle \hat{B}(t) \hat{B}(0) \right\rangle - \Braket{\hat{B}(t)} \Braket{\hat{B}(0)},
	\label{eqConnCorr}
	\end{align}
	where $\hat{B}(t)$ is an observable in the Heisenberg picture, and the average is taken with respect to some reference initial density matrix $\hat{\rho}_0$ (which may or may not be mixed, but is assumed to be stationary $[\hat{H},\hat{\rho}_0] = 0$). The Fourier transform of $\Gamma_B(t)$,
	\begin{align}
	\Gamma_B(\omega) \coloneqq \int \diff t\, e^{\iu \omega t} \Gamma_B(t),
	\label{eqSpecFun}
	\end{align}
	is the spectral function of $\hat{B}$. In the absence of external time-dependence, the spectral function can be written
	\begin{align}
	\Gamma_B(\omega) = \sum_{n \neq m} p_n \left\vert\braket{m|\hat{B}|n}\right\vert^2 \delta(\epsilon_m - \epsilon_n - \omega),
	\end{align}
	where $p_n = \braket{n|\hat{\rho}_0|n}$ are classical probabilities, and the sums are over eigenstates. A peak in the spectral function at a frequency $\omega$ indicates the existence of eigenstates which differ by an energy $\omega$.
	
	In order to be sensitive to the edge mode degeneracy, $\hat{B}$ must be charged under some of the symmetries, such that $\braket{1|\hat{B}|0}$ is non-zero, and the initial state should have some weight in the edge mode subspace. In this case, $\Gamma_B(\omega)$ exhibits a characteristic sharp peak at $\omega = 0$. We therefore take $\hat{B}$ to have non-zero $\mathbb{Z}_2$ charge under the symmetry corresponding to $\hat{\tau}^z$.  When working with spin chains, $\hat{B}$ is typically a magnetic field which is odd under the relevant spin-flip operator.
	
	We now subject the system to classical noise. The total Hamiltonian $\hat{H}(t)$ is then the sum of the unperturbed part $\hat{H}_0$ and a noise Hamiltonian $\hat{V}(t)$ which can be decomposed into an arbitrary number $M$ of independent channels, each varying stochastically:
	\begin{align}
	\hat{V}(t) = \sum_{\alpha=1}^M \eta_\alpha(t) \hat{V}_\alpha.
	\label{eqNoiseDecomp}
	\end{align}
	All noise operators $\hat{V}_\alpha$ are Hermitian, and respect all the relevant symmetries of the system. The signals $\eta_\alpha(t)$ are real, have zero mean, and are mutually uncorrelated. We use Gaussian noise, which can be completely characterized by the second moment
	\begin{align}
	\overline{\eta_\alpha(t) \eta_\beta(t')} = \delta_{\alpha \beta} \tilde{S}_\alpha(t-t'),
	\end{align}
	where $\tilde{S}_\alpha(t-t')$ is the Fourier transform of the noise spectral function $S_\alpha(\omega)$. Overlines denote averages over noise realisations. The noise is chosen to be low-frequency, such that the noise spectrum $S_\alpha(\omega)$ has no weight above the bulk energy gap $E_\text{g}$. This ensures that transitions in or out of the ground state subspace are energetically forbidden. We take the noise correlator to be invariant under the reversal of time $\tilde{S}_\alpha(t) = \tilde{S}_\alpha(-t)$, so that time-reversal symmetry is satisfied on average. We emphasise that the noise instantaneously respects all the symmetries, and there is no bias of the forward or backward direction of time -- this noise could therefore be achieved by coupling to a classical bath which is itself time-reversal symmetric.
	
	The calculation of the two-time correlation function \eqref{eqConnCorr} in the presence of low-frequency noise can be carried out using an adiabatic ansatz for the time-dependence of the wavefunction, since transitions out of the ground state subspace are energetically forbidden. Details of this calculation can be found in Appendix \ref{appNoise}. The evolution can be separated into a dynamical phase and a geometric (Berry) phase, the former of which has no observable effect, since the two ground states are always degenerate. For a particular realisation of the noise, we write the instantaneous ground states of $\hat{H}_0 + \hat{V}(t)$ as $\ket{0_{\hat{V}(t)}}$, $\ket{1_{\hat{V}(t)}}$, through which the correlator can be expressed as (taking only the terms that contribute to the spectral function at low frequencies)
	\begin{align}
	\Gamma_B(t) &= |B_{01}|^2 \braket{0|0_{\hat{V}(t)}} \braket{1_{\hat{V}(t)}|1} \nonumber\\ &\times\exp\left( -\int^t_0 \diff t'\, [\mathcal{A}_{11}(t') - \mathcal{A}_{00}(t')] \right),
	\label{eqExpConnectionMain}
	\end{align}
	where $B_{01} \coloneqq \braket{1|\hat{B}|0}$, and $\mathcal{A}_{11}(t) \coloneqq \braket{1_{\hat{V}(t)} | (\diff/\diff t) |  1_{\hat{V}(t)}}$ is the Abelian Berry connection (and similar for $\mathcal{A}_{00}(t)$).
	
	The symmetries of the system naturally constrain the form of the geometric phases induced by the noise. We show in Appendix \ref{appNoise} that in the case of a unitary $\mathbb{Z}_2 \times \mathbb{Z}_2$ symmetry, the Berry connections associated with each ground state are identical, provided a symmetry-respecting gauge is chosen
	\begin{align}
	\mathcal{A}_{11}(t) &= \mathcal{A}_{00}(t). &(\mathbb{Z}_2 \times \mathbb{Z}_2)
	\label{eqConnSymUnitMain}
	\end{align}
	In this case, the phases in the exponential of \eqref{eqExpConnectionMain} cancel, and the zero-frequency peak in the spectral function remains sharp. (The overlaps $\braket{0|0_{\hat{V}(t)}}$ are treated in Appendix \ref{appNoise}, and do not affect this conclusion.)
	
	On the other hand, when the phase is protected by the group $\mathbb{Z}_2 \times \mathbb{Z}_2^T$, the Berry connections for the two ground states satisfy
	\begin{align}
	\mathcal{A}_{11}(t) &= -\mathcal{A}_{00}(t), &(\mathbb{Z}_2 \times \mathbb{Z}_2^T)
	\label{eqConnSymAntiuMain}
	\end{align}
	which is natural since complex phases such as the Berry phase are not invariant under the action of antiunitary symmetries. The geometric phases induced by the noise therefore do not cancel, and accumulate in time. After averaging over noise realizations, the correlator is given by
	\begin{align}
	\overline{\Gamma_B(t)} = |B_{01}|^2 \exp\left(-4\overline{\theta_B(t)^2}\right)
	\end{align}
	where $\theta_B(t)$ is the total Berry phase after a time $t$. Although the statistical time-reversal symmetry $\tilde{S}_\alpha(t) = \tilde{S}_\alpha(-t)$ ensures that the Berry phase is zero on average ($\overline{\theta_B(t)} = 0$), each noise realization induces some non-zero phase, so that $\overline{\theta_B(t)^2}$ is non-zero. 
	The result is a spectral function of the form
	\begin{align}
	\overline{\Gamma_B(\omega)} = |B_{01}| \frac{\gamma}{\omega^2 + \gamma^2}
	\label{eqSpecBroad}
	\end{align}
	where the width of the spectroscopic peak $\gamma$ scales as
	\begin{align}
	\gamma \sim \frac{V^4}{\tau_n E_g^4},
	\label{eqScaling}
	\end{align}
	where $V$ is a characteristic strength of the noise signal, $\tau_n$ is the noise correlation time, and $E_g$ is the bulk energy gap. (See Appendix \ref{appNoise} for a derivation.)
	
	We see that antiunitary symmetries do not protect against decoherence between the degenerate ground states, leading to a broadening of the zero-energy peak with a width that scales according to Eq.~\eqref{eqScaling}. For a more general symmetry group $G$, we expect that the peak will remain sharp only if the edge degeneracy can be protected by the unitary symmetries alone -- this is equivalent to the condition for a system to be non-trivial in the non-equilibrium classification.
	
	We again emphasise that the noise we consider here instantaneously respects all symmetries, and does not break time-reversal symmetry on average, i.e.~the noise correlators are invariant under the reversal of time $\tilde{S}_{\alpha}(t) = \tilde{S}_{\alpha}(-t)$. In a related study, the authors of Ref.~\cite{Gao2016} also considered noise of this type, and correctly showed that this forces certain scattering amplitudes to be zero on average. However, here we show that this does not imply the lack of decoherence: Although the phase of the quantity in question (in our case, a geometric phase) is zero on average, each noise realization induces a particular phase which results in a non-zero rate of dephasing.

	\subsection{Application to Rydberg atom Experiment \label{secRyd}}
	
	In the above, we demonstrated that systems which are trivial in the non-equilibrium classification are susceptible to low-frequency noise, in that the characteristic zero-energy peak in spectroscopy will generally be broadened. Here, we explain how these effects could be observed in the context of a recent experiment using Rydberg atoms to realise an interacting bosonic SPT phase \cite{Leseleuc2018}.
	
	The experiment in Ref.~\cite{Leseleuc2018} realises an effective hopping model for hard-core bosons $\hat{b}_j$, $\hat{b}_j^\dagger$ satisfying $[\hat{b}_j, \hat{b}_{j'}^\dagger] = \delta_{jj'}[1 - 2\hat{b}_j^\dagger \hat{b}_j]$. Hopping processes between nearest-neighbours are implemented with an amplitude that is staggered on alternating bonds. One can represent the system as a spin-1/2 chain spanned by Pauli matrices $\hat{\sigma}^x_j = \hat{b}_j + \hat{b}^\dagger_j$, $\hat{\sigma}^z_j = (2\hat{b}_j^\dagger\hat{b}_j - 1)$, whereby the Hamiltonian can be written as
	\begin{align}
	\hat{H} = \sum_{j=1}^{N-1} \left[J - (-1)^j \delta \right]\left(\hat{\sigma}^x_j \hat{\sigma}^x_{j+1} + \hat{\sigma}^y_j\hat{\sigma}^y_{j+1}\right)
	\end{align}
	where $J$ is the average hopping amplitude and $\delta$ is the strength of the staggering.
	
	A Jordan-Wigner transform reveals that this system is in the non-trivial phase of the SSH model when $\delta < 0$, however the protecting symmetries in the bosonic model are different. Overall, the symmetry group is $(U(1) \rtimes \mathbb{Z}_2) \times \mathbb{Z}_2^T$, where the three factor groups are respectively generated by the three operators
	\begin{subequations}
		\begin{align}
		\hat{U}_\theta &= \exp \left(i\theta\sum\nolimits_j \hat{\sigma}^z \right) \label{eqSpin} \\
		\hat{C} &= \prod_j \hat{\sigma}^x \label{eqFlip} \\
		\hat{T} &= \prod_j \left( \iu\hat{\sigma}^y\right) \hat{\mathcal{K}} \label{eqTRS}
		\end{align}
	\end{subequations}
	Note that only the last of these three generators is antiunitary. All generators commute up to an unphysical phase with the exception of $\hat{C} \hat{U}_\theta = \hat{U}_{-\theta} \hat{C}$.
	
	In the non-trivial phase, an effective spin-1/2 degree of freedom appears at each edge, with a degeneracy protected by the above symmetries. This gapless degree of freedom was identified in the experiment using microwave spectroscopy \cite{Leseleuc2018}.
	One can show that the unitary symmetries alone [\eqref{eqSpin} and \eqref{eqFlip}] are sufficient to protect this degeneracy, and so the phase is non-trivial in the non-equilibrium classification. We therefore expect this spectroscopy peak to remain sharp in the presence of low-frequency noise.
	
	However, one can engineer extra terms in the Hamiltonian which break some of the symmetries such that the system is still in a topological phase. The full symmetry group has various subgroups, which include the $\mathbb{Z}_2 \times \mathbb{Z}_2$ and $\mathbb{Z}_2 \times \mathbb{Z}_2^T$ groups we considered in Section \ref{secNoise}. If we add a term of the form $\hat{\sigma}^x_j \hat{\sigma}^y_k$, then only the second symmetry subgroup is respected, generated by $\hat{U}_\pi$ and $\hat{T}$. The action of these symmetries on the edge mode is the same as in Eq.~\eqref{eqSymOps}. From the previous section, we know that systems protected in this way will exhibit broadened spectral peaks when low-frequency noise is present.
	
	One could therefore verify our predictions in this experiment by comparing the width of the zero-frequency spectral peak in the presence of noise when the perturbing term $\hat{\sigma}^x_j \hat{\sigma}^y_k$ is absent or present. One would only expect to see noise-induced broadening when this term is applied, since only then is the system trivial in the non-equilibrium classification. 
	
	\subsection{General Applicability of the Classification}
	
	We have demonstrated, by means of examples, that our non-equilibrium classification can be use to understand the topological aspects of non-equilibrium systems in a variety of scenarios. Let us briefly discuss the general conditions under which our results can be of use.
	
	Firstly, our classification scheme can be applied independently of the way in which the system is driven out of equilibrium. If the details of $\hat{H}(t)$ contain some additional structure (e.g.~if the Hamiltonian is time-periodic), then it is possible that topological effects can arise that are protected by this structure, e.g.~in Floquet-SPT systems \cite{vonKeyserlingk2016,Potter2016,Else2016}. In these cases, a different classification scheme should be applied that is appropriate to the specific type of dynamics. However, the non-equilibrium classification represents the most general scheme that can make predictions for generic driving protocols.
	
	Secondly, the results apply in regimes where the relevant wavefunctions remain area-law entangled (due to the condition of `finite-time' evolution). The typical example of when this occurs is after a quench from some other area-law entangled state. There, the classification will capture the topological properties of $\ket{\Psi(t)}$ for times less than some critical time that is extensive in the system size; beyond this time, the system is expected to be in the approach to some equilibrium state, which is not captured by our classification. However, as we saw in the case of classical noise, it is also relevant when the system is not driven \textit{far} from equilibrium. For low-frequency drives (below the bulk gap), the bulk is unperturbed, and $\ket{\Psi(t)}$ remains area-law entangled for much longer times. As long as $\hat{H}(t)$ is always instantaneously within the same equilibrium phase, we expect that the dynamics of the topological edge modes will modified in a way that reflects the results of the non-equilibrium classification.
	
	If one wishes to construct a classification that applies to non-equilibrium scenarios that feature volume-law states, then we expect that a different constraint will need to be imposed on the wavefunctions that are being considered (since the space of \textit{all} symmetry-respecting states in a Hilbert space is topologically trivial). An understanding of the different types of states that are relevant to other regimes of non-equilibrium dynamics could help to motivate future studies beyond our work.
	
	More generally, the non-equilibrium classification captures an important difference between the r{\^o}les of symmetries in equilibrium and non-equilibrium dynamics. While antiunitary symmetries are capable of protecting topological phenomena in equilibrium scenarios, they are no longer able to do so in generic non-equilibrium scenarios. This is reflected in the structure of Tables \ref{tab1D} and \ref{tabAllDim}.

	\section{Conclusions\label{secConc}}
	
	We have provided a construction for topologically classifying many-body wavefunctions in a way which naturally pertains to non-equilibrium scenarios. The classification scheme applies to short-ranged entangled wavefunctions protected by some symmetry group.  Thus, much like its equilibrium counterpart (Ref.~\onlinecite{Chen2010}), it naturally incorporates both weakly and strongly correlated systems. It differs from the equilibrium classification in the definition of how one is permitted to deform one wavefunction into another. Wavefunctions are considered as topologically equivalent if one can be deformed to the other through some finite-time unitary evolution under a Hamiltonian which respects the specified symmetries. The finite time of evolution ensures that the wavefunctions remain short-ranged entangled, but the possibility of breaking antiunitary symmetries along this trajectory allows for a wider space of intermediate states to be explored, and so topologically distinct states in equilibrium may be indistinct out of equilibrium. The equivalence classes under this relation constitute the non-equilibrium classification [Fig.~\ref{figClass}], which is summarised in Tables \ref{tab1D} and \ref{tabAllDim}. The first table was derived by considering how one-dimensional wavefunctions projectively represent the governing symmetry transformations before and after a quench. The second table generalises these methods to higher dimensions, making use of the relationship between SPT states and group cohomology. We then demonstrated how our findings predict universal phenomenology in various non-equilibrium scenarios, focussing on the effect of external noise on the spectroscopy of SPT systems.

	The previously established results for free-fermion systems out of equilibrium \cite{McGinley2018,McGinley2019} can be reproduced when the above construction is applied to that context. We therefore expect that many of the experimental signatures associated with those non-interacting systems will also apply to the interacting case, including the dynamics of bulk topological indices \cite{McGinley2018}, and the decoherence of quantum information stored in topological bound states \cite{McGinley2019}. Given the increasing number of non-interacting and interacting SPT phases which are accessible in ultracold atom and other high-precision experiments, and the long coherence times that they promise, we expect that these signatures should be observable with currently used technologies and techniques.
	
	We note that in the context of open systems with Markovian environments, a classification scheme for density matrices has recently been suggested which bears a resemblance to our pure-state construction: Density matrices can be classified according to whether one can evolve from one to the other in a finite time using a Lindblad master equation \cite{Coser2018}. As in our scheme, the condition of evolving for a finite time prohibits the spread of correlations over arbitrary distances, playing the same role as the bulk gap in the equilibrium classification. However, the possibility of entangling the system with the environment allows for an even more general deformation procedure than the finite-time unitary evolution. Given the relevance of our classification to the effect of noise on experimental systems (albeit low-frequency noise), it will be interesting to understand how these approaches are related in future work.
	
	\begin{acknowledgements}
	This work was supported by an EPSRC studentship and
	Grant Nos.~EP/P034616/1 and EP/P009565/1, and by an Investigator Award of the Simons Foundation. Statement of compliance with EPSRC policy framework on research data: All data are directly available within the publication.
	\end{acknowledgements}
	
	\appendix

	\section{Computing the non-equilibrium classification using the Hochschild-Serre spectral sequence\label{secSS}}
	
	In this appendix, we demonstrate how to compute the non-equilibrium classification, which in Section \ref{secHigher} was shown to be the image of the restriction functor from $\mathcal{H}^{d+1}[G,U_T(1)]$ to $\mathcal{H}^{d+1}[G_T,U_T(1)]$, where $G_T$ is the subgroup of $G$ containing only unitary elements [Eq.~\eqref{eqClass}]. The restriction functor is a well-studied object in group cohomology, and it features in a theorem known as the Hochschild-Serre (HS) spectral sequence (sometimes referred to as the Lyndon spectral sequence). We provide a brief discussion of spectral sequences, and how they apply to our classification problem; however for a more formal introduction to the methods used, see e.g.~Ref.~\cite{Brown1982}.
	
	A spectral sequence is best visualised on a three dimensional grid. Each 2D layer is referred to as a \textit{page} (or sometimes leaf, sheet, or term), labelled by $r \geq 0$. There is some initial page $r_0$, which is often $r = 1$ or $2$. On each page, there is a square grid, and each point $(p,q)$ of the grid, there is a group, denoted $E_r^{p,q}$. We have $E_r^{p,q} = \{e\}$, the trivial group, for $p < 0$ or $q < 0$. The relationships between different pages are constructed through the \textit{differentials} $d_r^{p,q} : E_r^{p,q} \rightarrow E_r^{p+r,q-r+1}$. Note here that within the page $r$, the domain and codomain of the differentials are relatively displaced by a vector which depends on $r$. The differentials satisfy $d_r^{p+r,q-r+1} \circ d_r^{p,q} = 0$, which means that one can form cochain complexes between the groups $E_r^{p+nr,q-nr+1}$, where $n$ runs over the integers.
	
	The groups which constitute the $(r+1)$th page are the cohomology groups of the cochain complexes within the $r$th page, i.e.~
	\begin{align}
	E_{r+1}^{p,q} \cong \ker d_r^{p,q} / \im d_r^{p-r,q+r-1}
	\end{align}
	The sequence continues for increasing page number $r\rightarrow \infty$. In words, spectral sequences are built recursively by constructing cochain complexes between cohomology groups of the previous cochain complexes.
	
	We say that a spectral sequence converges if for $r$ large enough, the groups $E_r^{p,q}$ become independent of $r$. For example, $E_r^{1,1}$ must converge for $r \geq 3$, since $d_3^{1,1}$ has codomain $E_3^{4,-1} = 0$ and $d_3^{-2,3}$ has domain $E_3^{-2,3} = 0$, hence $E_4^{1,1} = E_3^{1,1}$, and so on. One denotes this convergence as
	\begin{align}
	E_r^{p,q} \Rightarrow H^{p+q}.
	\end{align}
	Here the groups $H^{p+q}$ are \textit{not} given by $E_{\infty}^{p,q}$, but have structure related to the converged groups. Specifically, there is a filtration
	\begin{align}
	0=H_{n+1}^n \subseteq H_n^n \subseteq H_{n-1}^n \subseteq \cdots \subseteq H_{1}^n \subseteq H_0^n = H^n
	\label{eqFilt1}
	\end{align}
	such that
	\begin{align}
	E_\infty^{p,n-p} \cong H^n_p/H^n_{p+1}
	\label{eqFilt2}
	\end{align}
	Thus the diagonals on page $r \rightarrow \infty$ with fixed $p+q$ are quotients of successive subgroups of $H^{p+q}$. Knowledge of the $E_\infty^{p,q}$ thus provides a lot of information about $H^{p+q}$. Two particularly simple data are the edge maps
	\begin{subequations}
		\begin{align}
		E_{r_0}^{n,0} \twoheadrightarrow &E_{\infty}^{n,0} \hookrightarrow H^{n} \label{eqEdgeMapsA} \\
		H^n  \twoheadrightarrow &E_{\infty}^{0,n} \hookrightarrow E_{r_0}^{0,n},
		\label{eqEdgeMapsB}
		\end{align}
	\end{subequations}
	The surjectivity (indicated by $\twoheadrightarrow$) of the first map of each row follows since $E_{\infty}^{n,0}$ is a quotient of $E_{r_0}^{n,0}$, and $E_\infty^{0,n} \cong H^n/H_1^n$ is a quotient of $H^n$. Injectivity (indicated by $\hookrightarrow$) follows since $E_\infty^{n,0} \cong H_n^n$ is a subgroup of $H^n$, and $E_{\infty}^{0,n}$ is a subgroup of $E_{r_0}^{0,n}$.
	
	The Hochschild-Serre spectral sequence relates cohomology groups of $G$ to cohomology groups of one of its subgroups. Specifically, if $H$ is a normal subgroup of $G$, then it is expressed as
	\begin{align}
	E_2^{p,q} = \mathcal{H}^{p}(G/H,\mathcal{H}^q(H,M)) \Rightarrow \mathcal{H}^{p+q}(G,M)
	\label{eqHSSpec}
	\end{align}
	where $M$ is an arbitrary $G$-module. The usage of $\mathcal{H}^q(H,M)$ as a $G/H$-module means we must understand how elements of $G/H$ act on elements of the cohomology group, via the action of $H$ on $M$. Examples of this construction can be found in Appendix J.10 of Ref.~\cite{Chen2013}.
	
	The HS spectral sequence is usually applied as a tool for computing cohomology for complicated groups $G$ based on some simpler structure contained in its subgroups. Here, instead, we assume that $\mathcal{H}^{p+q}(G,M)$ is already known by some other method, and make use an important corollary, namely that the composite edge map in Eq.~\eqref{eqEdgeMapsB} is given by the restriction functor from $H^n = \mathcal{H}^{n}(G,M)$ to $\mathcal{H}^{n}(H,M)$ \cite{Weston}. (Note that $E_{r_0}^{0,n} = \mathcal{H}^{0}(G/H,\mathcal{H}^n(H,M)) = \mathcal{H}^n(H,M)^{G/H} \leq \mathcal{H}^n(H,M)$, where the superscript $A^{G/H}$ denotes the submodule of $A$ which is invariant under the action of elements of $G/H$.) Given the respective surjectivity and injectivity of the left and right maps in \eqref{eqEdgeMapsB}, we have
	\begin{align}
	\im \text{Res}_n \cong E_{\infty}^{0,n},
	\end{align}
	the left hand side of which gives the non-equilibrium classification in dimension $d = n-1$, by the arguments of Section \ref{secHigher}. Our task is now reduced to finding how the HS spectral sequence \eqref{eqHSSpec} converges along the $(0,n)$ axis, using $H = G_T$ as the subgroup, and $M = U_T(1)$ as the $G$-module. We can also simplify matters further by using the isomorphism $\mathcal{H}^n(G,U_T(1)) \cong \mathcal{H}^{n+1}(G,\mathbb{Z}_T)$, where $\mathbb{Z}_T$ is a $G$-module with underlying group $\mathbb{Z}$, transforming as $T : a \mapsto -a$ for antiunitary elements $T$.
	
	We now describe how to obtain $E_{\infty}^{0,n}$ for the group $G = \mathbb{Z}_n \times \mathbb{Z}_m \times \mathbb{Z}_2^T$, module $M = \mathbb{Z}_T$, and $n = 2,\ldots,5$ corresponding to spatial dimension $d = 0,\ldots,3$. The other groups in Table \ref{tabAllDim} can be obtained in similar ways.
	
	The second page of the sequence $E_2^{p,q} = \mathcal{H}^p(\mathbb{Z}_2^T,\mathcal{H}^q(\mathbb{Z}_n \times \mathbb{Z}_m,\mathbb{Z}_T))$ can be calculated using identities for cohomologies of finite cyclic groups \cite{Joyner2007}, giving
	\begin{widetext}
		\begin{equation}
		E_2^{p,q} \hspace{5pt} = \hspace{10pt} \begin{array}{cc|cccccc}
		& 0 & B_{n,m} \times B_{n,m} & B_{n,m} \times B_{n,m} & B_{n,m} \times B_{n,m} & B_{n,m} \times B_{n,m} & B_{n,m} \times B_{n,m} & B_{n,m} \times B_{n,m} \\
		& 0 & A_{n,m} \times B_{n,m} & A_{n,m} \times B_{n,m} & A_{n,m} \times B_{n,m} & A_{n,m} \times B_{n,m} & A_{n,m} \times B_{n,m} & A_{n,m} \times B_{n,m} \\
		\uparrow & 0 & B_{n,m} & B_{n,m} & B_{n,m} & B_{n,m} & B_{n,m} & B_{n,m} \\
		q& 0 & A_{n,m} & A_{n,m} & A_{n,m} & A_{n,m} & A_{n,m} & A_{n,m} \\
		& 0 & 0 & 0 & 0 & 0 & 0 & 0 \\
		& 0 & 0 & \mathbb{Z}_2 & 0 & \mathbb{Z}_2 & 0 & \mathbb{Z}_2\\ \hline
		& 0 & 0 & 0 & 0 & 0 & 0 & 0 \\
		& & &  & p \rightarrow & & &
		\end{array}
		\label{eqPage2}
		\end{equation}
	\end{widetext}
	where we have used the shorthand
	\begin{align}
	A_{n,m} &\coloneqq \mathbb{Z}_{(2,n)} \times \mathbb{Z}_{(2,m)} \nonumber\\
	B_{n,m} &\coloneqq \mathbb{Z}_{(2,n,m)}
	\end{align}
	and $(a,b,c,\ldots)$ denotes the greatest common divisor of the integers in the brackets. The converged page $E_{\infty}^{p,q}$ is obtained by applying the differentials $d_r^{p,q}$ sequentially, however there is generally no explicit expression for these $d_r^{p,q}$. Instead, we can note from Eqs.~\eqref{eqFilt1}, \eqref{eqFilt2}, that the diagonals of $E_{\infty}^{p,q}$ with $p+q = n$ provide a filtration of $\mathcal{H}^n(\mathbb{Z}_n \times \mathbb{Z}_m \times \mathbb{Z}_2^T,\mathbb{Z}_T)$, which itself was calculated by independent means using the torsion product in Appendix J.7 of Ref.~\onlinecite{Chen2013}. For example, when $n = 5$ (corresponding to 3 spatial dimensions), we have $\mathcal{H}^5(\mathbb{Z}_n \times \mathbb{Z}_m \times \mathbb{Z}_2,\mathbb{Z}_T) = \mathbb{Z}_2 \times B_{n,m}^{\times4} \times A_{n,m}^{\times2}$.
	
	Now, with increasing $r$, the group $E_{r+1}^{p,q} = \ker d_r^{p,q} / \im d_r^{p-r,q+r-1}$ can either remain the same as $E_{r}^{p,q}$ (if $d_r^{p,q}$ and $d_r^{p-r,q+r-1}$ are both the zero map), or become a smaller group. However, we see that already at the $r=2$ page, the product of the orders of the groups is $\prod_p |E_2^{p,5-p}| = |\mathbb{Z}_2| |B_{n,m}|^4 | A_{n,m}|^2$, which matches the order of the previously obtained cohomology group  $=|\mathcal{H}^5(\mathbb{Z}_n \times \mathbb{Z}_m \times \mathbb{Z}_2,\mathbb{Z}_T)|$. We also know from the filtration Eqs.~\eqref{eqFilt1}, \eqref{eqFilt2} that $|\mathcal{H}^5(\mathbb{Z}_n \times \mathbb{Z}_m \times \mathbb{Z}_2,\mathbb{Z}_T)| = \prod_p |E_\infty^{p,5-p}|$. Given that $|E_\infty^{p,q}| \leq |E_2^{p,q}|$, with equality implying $E_\infty^{p,q} = E_2^{p,q}$, we conclude that along this diagonal, the sequence already converges at the 2nd page. The same turns out to be true for all the diagonals $p+q = 2,\ldots,5$. Therefore, in this case, the non-equilibrium classification equals $E_2^{0,d+2}$, given by the first non-trivial column of \eqref{eqPage2}.
	
	The remaining symmetry classes can be obtained using the same method. However, some follow more easily: if the only homomorphism from $\mathcal{H}^{1+d}[G,U_T(1)]$ to $\mathcal{H}^{1+d}[G_T,U_T(1)]$ is the trivial map, then its image must be trivial. For example in the case of $G = U(1) \rtimes \mathbb{Z}_2^T$ in $d = 2$ dimensions, there is no non-trivial homomorphism from $\mathbb{Z}_2$ to $\mathbb{Z}$.
	
	\section{Examples in 2D\label{secExamples}}
	
	Here we provide some concrete examples of non-equilibrium protocols (specifically quenches) which exhibit the physics captured by the non-equilibrium classification (Table \ref{tabAllDim}), and confirm the results which we found using the cohomology methods of Section \ref{secHigher}. In particular, we will consider two-dimensional systems with the symmetry group $\mathbb{Z}_2 \times \mathbb{Z}_2^T$, which has a classification $\mathbb{Z}_2 \times \mathbb{Z}_2 \rightarrow \mathbb{Z}_2$ (equilibrium $\rightarrow$ non-equilibrium). In this section, we write $a$ for the generator of the unitary $\mathbb{Z}_2$ subgroup, and $T$ for the generator of the antiunitary $\mathbb{Z}_2^T$ subgroup.

	\begin{figure}
		\includegraphics[scale=1]{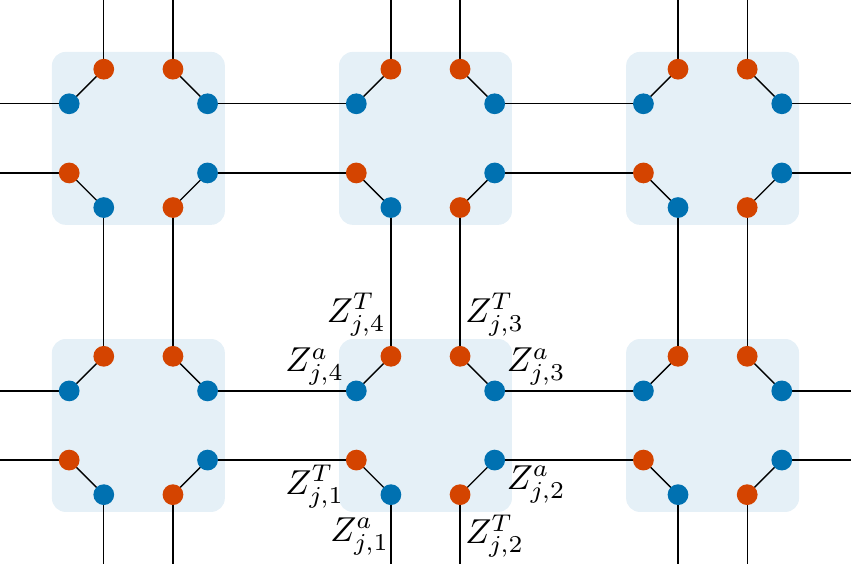}
		\caption{Illustration of the ground state of Hamiltonian $H_0$ [Eq.~\eqref{eqHamProj}]. Each physical site (light blue squares) hosts eight spins, which are acted on by the Pauli matrices $Z_{j,m}^f$. Spins with $a$ flavour are dark blue, and spins with $T$ flavour are red. In the ground state, the two spins on a site with the same $m$ index are entangled with six other spins on neighbouring sites, as shown by the solid lines. Each octet of spins are in the state $\ket{\Phi_p}$ [Eq.~\eqref{eqPlaquette}]. To reveal the SPT order of the ground state, the system is divided along the dashed line, and the entanglement spectrum of the reduced density matrix for the bottom half is studied (see main text).}
		\label{figPlaquettes}
	\end{figure}

Our results are easiest captured using exactly solvable models, which we construct in an analogous way to the $\mathbb{Z}_2$-symmetric CZX model of Ref.~\cite{Chen2011b}. For the models we use, the Hilbert space on each site is made up of eight spin-1/2s. Within a site, spins are labelled by a subcell-position index $m = 1,\ldots,4$, and a `flavour' index $f = a, T$, which are associated with the two generators of the symmetry group. The bosonic operators are generated by Pauli matrices for each spin $X^f_{j,m}, Y^f_{j,m}, Z^f_{j,m}$, where $j$ labels the lattice site.
	
	As explained in Ref.~\cite{Chen2013}, an exactly solvable SPT phase can be constructed through an appropriate choice of the symmetry group action on the local Hilbert space, and by choosing a Hamiltonian such that each spin is entangled in a plaquette pattern. The two models constructed in this section have the same initial Hamiltonian, but the symmetry operations are defined differently for each. The pattern of entanglement generated by this Hamiltonian is illustrated in Figure \ref{figPlaquettes}.  The two spins with a given $m = 1,\ldots,4$ couple to other spin-pairs on the four sites of the relevant plaquette. Specifically, the Hamiltonian takes the form
	\begin{align}
	H_0 = -\sum_p X^{(8)}_p \otimes P^{(4)}_{p,u} \otimes P^{(4)}_{p,d}\otimes P^{(4)}_{p,l}\otimes P^{(4)}_{p,r}
	\label{eqHamProj}
	\end{align}
	where the sum is over all plaquettes $p$. The factor $X^{(8)}_p$ acts on all eight spins involved in the plaquette as $X^{(8)}_p = \ket{\Phi_p}\bra{\Phi_p}$, where
	\begin{align}
	\ket{\Phi_p} = \sum_{\substack{n_a=\pm \\ n_T=\pm}} \ket{n_a}_{p,1a}\otimes \ket{n_T}_{p,1T} \otimes \cdots \otimes \ket{n_a}_{p,4a}\otimes \ket{n_T}_{p,4T}
	\label{eqPlaquette}
	\end{align}
	where $p, 1a$ denotes the spin with flavour $a$ and subcell index 1, belonging to plaquette $p$ (see Fig. \ref{figPlaquettes}). The projectors $P^{(4)}_{p,u}$ are defined on the four spins directly above ($u$), below ($d$), left ($l$),
	or right ($r$) of the plaquette $p$. These take the form
	\begin{align}
	P^{(4)}_{p,u} = (1 - Z_{p\uparrow,3}^a Z_{p\uparrow,4}^a)(1 - Z_{p\uparrow,3}^T Z_{p\uparrow,4}^T)
	\end{align}
	and similar, where $p\uparrow$ denotes the plaquette above $p$. All terms defined so far commute, and project onto the unique ground state $\ket{\Psi_0}=\bigotimes_p \ket{\Phi_p}$.
	
	The above Hamiltonian can only be identified as being in an SPT phase once we define the symmetry action on the Hilbert space. As in the main text, we only consider on-site symmetry action, but since four plaquettes overlap on one site, the symmetry operation can have non-trivial action between entangled degrees of freedom. We consider our two models sequentially.
	
	\emph{Model 1.---} The first model realizes the $\mathbb{Z}_2 \times \mathbb{Z}_2^T$ symmetry through acting on a site $j$ in the following way
	\begin{align}
	U(a)_j &= CZ^a_{j,12}CZ^a_{j,23}CZ^a_{j,14}CZ^a_{j,43} \prod_{m=1}^4 X_{j,m}^a Z_{k,m}^T \nonumber\\
	U(T)_j\mathcal{K} &= \prod_{m=1}^4 X_{j,m}^T \mathcal{K}.
	\label{eqModel1}
	\end{align}
	Here, in analogy to Ref.~\onlinecite{Chen2011b}, $CZ^a_{j,mn}$ is the controlled-$Z$ operator on the two spins $Z_{j,m}^a$ and $Z_{j,n}^a$:
	\begin{align}
	CZ^a_{j,mn} = \frac{1}{4}(1 + Z_{j,m}^a)(1 + Z_{j,n}^a),
	\end{align}
	and $\mathcal{K}$, which represents complex conjugation, effects the antiunitary action of the generator $T$. 
	
	The SPT order of the ground state $\ket{\Psi_0}$ is easiest seen through its entanglement spectrum. One partitions an infinite system into upper and lower halves ($A$ and $\bar{A}$), with boundary  between two rows of sites (dashed line in Fig.~\ref{figPlaquettes}). Tracing out the upper half results in a reduced density matrix $\rho_A = \Tr_{\bar{A}} \ket{\Psi_0}\bra{\Psi_0}$, whose eigenvalues constitute the entanglement spectrum \cite{Li2008}. The entanglement spectrum will be degenerate if $\ket{\Psi_0}$ has SPT-order \cite{Pollmann2010}.
	
	Since plaquettes are mutually unentangled, we can restrict ourselves to just one plaquette straddling the entanglement cut, and trace out the four spins located in $\bar{A}$. If one adds perturbations on top of $H_0$ \eqref{eqHamProj} which respect the symmetries \eqref{eqModel1}, then the entanglement degeneracies will persist as long as plaquettes arbitrarily far apart remain unentangled, i.e.~a critical point is not crossed. For example, the term $Y^a_{j,3}Y^a_{j,1}$, where $j$ is a site just above the entanglement cut, is allowed by symmetry. Since this term only entangles two plaquettes, we can still find the ground state using exact diagonalization techniques on 16 spins. We find that the entanglement spectrum remains degenerate when this term, and indeed any other symmetry-respecting term, is added to $H_0$. This verifies that, in equilibrium, this model is in an SPT phase.
	
	We now describe a non-equilibrium protocol which only couples a finite number of plaquettes, and thus is amenable to exact diagonalization. The initial state at time $t=0$ is the unperturbed ground state $\ket{\Psi_0}$. We then time evolve under Hamiltonian $H_1+H_2(t)$. The first part $H_1$ contains two of the terms featuring in $H_0$ [Eq.~$\eqref{eqHamProj}$], featuring one plaquette $p_1$ which crosses the entanglement cut and another $p_2$ which is one plaquette up and to the right of $p_1$, such that they have one physical site in common. The second part $H_2(t)$ will contain various perturbations which couple plaquettes $p_1$ and $p_2$, and may itself vary in time, but at any time $t$, $H_2(t)$ is invariant under the symmetries \eqref{eqModel1}. The resulting evolution can be simulated using 16 spins. We find, again, that so long as $H_1 + H_2(t)$ respects the symmetries, the time-evolved state still has entanglement degeneracies for arbitrary times. We conclude that in this model, topology is preserved out of equilibrium.
	
	\emph{Model 2.---} The second model we present is also an SPT phase in equilibrium, but exhibits very different phenomena out of equilibrium. In this case, the symmetries are
	\begin{align}
	U(a)_j &= \prod_{m=1}^4 X_{j,m}^a \nonumber\\
	U(T)_j\mathcal{K} &= CZ^T_{j,12}CZ^T_{j,23}CZ^T_{j,14}CZ^T_{j,43} \prod_{m=1}^4 X_{j,m}^T Z_{j,m}^a \mathcal{K},
	\label{eqModel2}
	\end{align}
	with all terms defined as above. Again, adding symmetry-respecting perturbations to $H_0$ does not remove the entanglement degeneracy of the ground state, and the model constitutes an SPT phase with the same protecting symmetry group as model 1 ($\mathbb{Z}_2 \times \mathbb{Z}_2^T$).
	
	Now, we consider the same non-equilibrium protocol as defined above. Let us consider the following piecewise form for $H_2(t)$ between times 0 and 3
	\begin{align}
	H_2(t) = \begin{cases}
	Y_{p_1,2}^aZ_{p_1,4}^a + \alpha Y_{p_1,1}^aZ_{p_1,3}^a & 0 < t \leq 1 \\
	X_{p_1,1}^T X_{p_2,3}^T & 1 < t \leq 2 \\
	Y_{p_1,1}^T X_{p_2,3}^T & 2 < t \leq 3
	\end{cases}
	\end{align}
	where $\alpha \neq 0$ is an arbitrary constant. All three terms satisfy the symmetries given by \eqref{eqModel2}. However, when we time-evolve under each Hamiltonian sequentially, we will find that the entanglement spectrum of the final state is non-degenerate, and hence topologically non-trivial. This behaviour is not specific to the choice of $H_2(t)$, but this piecewise evolution makes the mechanism clear, as we now describe.
	
	The Hamiltonian for $0 < t \leq 1$ has the effect of breaking the fourfold entanglement degeneracy of $\ket{\Psi_0}$ down to twofold; this would also occur for the ground state of a system with such a term included and thus is not a non-equilibrium effect. Now, the time evolution between $t = 1$ and $t = 3$ is governed by the unitary operator $\exp(-iX_{p_1,1}^T X_{p_2,3}^T)\exp(-iY_{p_1,1}^T X_{p_2,3}^T)$. This non-commuting product can be expressed as a single exponential using the Baker-Campbell-Hausdorff formula for $\log (e^A e^B)$, the leading terms of which are $A + B + [A,B]/2 + \cdots$. The commutator in this expression plays a crucial role: the series includes the term $-[X_{p_1,1}^T X_{p_2,3}^T,Y_{p_1,1}^T X_{p_2,3}^T]/2 = -iZ_{p_1,1}^T$. Thus, the evolution from $t = 1$ to $3$ proceeds as if it were generated by a static Hamiltonian with a term $\propto Z_{p_1,1}^T$. Such a term \textit{is not} invariant under $U(T)\mathcal{K}$. This manifestation of dynamically induced symmetry breaking leads to a non-degenerate entanglement spectrum after time evolution, and so we conclude that in this second model, topology is lost out of equilibrium.
	
	We have now provided two models, both with the same symmetry group $\mathbb{Z}_2 \times \mathbb{Z}_2^T$, one of which preserves its topology out of equilibrium, and the other of which does not. This confirms the non-equilibrium classification $\mathbb{Z}_2\times \mathbb{Z}_2 \rightarrow \mathbb{Z}_2$ for this group given in Table \ref{tabAllDim}, which we independently computed using cohomological methods.
	
	\section{Calculation of Spectral Function in the Presence of Noise \label{appNoise}}
	
	Here we provide details of the analytical calculations outlined in Section \ref{secNoise} regarding the spectral function \eqref{eqSpecFun} in the presence of low frequency noise, as specified in the main text. We are considering a one-dimensional system in an SPT phase with two degenerate ground states $\ket{0}$, $\ket{1}$, protected by either $\mathbb{Z}_2 \times \mathbb{Z}_2$ or $\mathbb{Z}_2 \times \mathbb{Z}_2^T$ symmetry groups.
	
	We start by calculating the connected correlator \eqref{eqConnCorr} in the time domain. For simplicity, we choose an initial state $\hat{\rho}_0 = \ket{0} \bra{0}$, although any mixed density matrix within the ground state subspace will exhibit the same spectral form. We also assume that $\hat{B}$ has definite non-zero charge under the symmetries, so that $\braket{\hat{B}(0)} = \braket{\hat{B}(t)} = 0$, and $\braket{1|\hat{B}|0} \neq 0$. (Most generally, however, $\hat{B}$ only need be non-invariant under the symmetry in order to see the zero energy peak.) For a single noise realization $\hat{V}(s)$ ($0 \leq s \leq t$), the two-time correlator in the interaction picture becomes
	\begin{align}
	\Gamma_B[\hat{V}(s)](t) = \Braket{ 0 | \hat{U}_I(t,0)^\dagger\, \hat{B}_I(t)\, \hat{U}_I(t,0)\, \hat{B}_I(0) |0},
	\end{align}
	where $\hat{U}_I(t,0)$ is the interaction-picture unitary evolution operator from time $0$ to $t$, which can be expressed as a time-ordered exponential. Expanding the interaction operator $\hat{B}(t)$ in the basis of eigenstates of the unperturbed SPT Hamiltonian $\hat{H}_0$, we find
	\begin{align}
	\hat{B}_I(t) = \sum_{nm}  \braket{m|\hat{B}|n} e^{\iu(\epsilon_m - \epsilon_n)t} \ket{m}\bra{n}.
	\end{align}
	The correlator can then be written
	\begin{align}
	\Gamma_B[\hat{V}(s)](t) &= \sum_{nmp} e^{\iu(\epsilon_m - \epsilon_n)t} \braket{m|\hat{B}|n} \braket{p|\hat{B}|0} \nonumber\\
	&\times  \braket{0|\hat{U}_I(t,0)^\dagger |m} \braket{n|\hat{U}_I(t,0) |p}.
	\label{eqSpectralDecomp}
	\end{align}
	We focus on low-frequency components of the spectral function $\omega \ll E_\text{g}$, where the edge mode peak is expected to be. This ensures that in the sum over eigenstates above, $\ket{n}$ and $\ket{m}$ must differ by an energy smaller than the bulk gap. Furthermore, as previously discussed, transitions into or out of the ground state subspace due to the noise are forbidden, i.e.~$\braket{0|\hat{U}_I(t,0)^\dagger|m}$ is non-zero only if $\ket{m}$ is a ground state. These energetic conditions restrict $\ket{n}$, $\ket{m}$, and $\ket{p}$ to each be either $\ket{0}$ or $\ket{1}$.
	
	We must also consider the symmetry properties of the eigenstates. The operator $\hat{B}$ is assumed to have odd $\mathbb{Z}_2$ charge under the $\hat{\tau}^z$ symmetry, so $\ket{m}$ and $\ket{n}$ must be in opposite charge sectors, and the same for $\ket{p}$ and $\ket{0}$. Furthermore, the time evolution operator $\hat{U}_I(t,0)$ is generated by the noise potential $\hat{V}(s)$, which itself respects a unitary $\mathbb{Z}_2$ symmetry in both symmetry groups chosen, so $[\hat{\tau}^z, \hat{U}_I(t,0)] = 0$ within the ground state subspace. Therefore $\ket{n}$ must have the same charge as $\ket{0}$, and $\ket{m}$ must have the same charge as $\ket{p}$. The energetic and symmetry restrictions on the matrix elements in \eqref{eqSpectralDecomp} leave only one non-zero term in the sum, namely
	\begin{align}
	\Gamma_B[\hat{V}(s)](t) &= |B_{10}|^2 \braket{0|\hat{U}_I(t,0)^\dagger |0} \braket{1|\hat{U}_I(t,0) |1} \nonumber\\
	&= |B_{10}|^2 \braket{0|\hat{U}(t,0)^\dagger |0} \braket{1|\hat{U}(t,0) |1}, \label{eqCorrRealization}
	\end{align}
	where $B_{10} = \braket{1|\hat{B}|0}$, and in the last equality we transform back to the Schr{\"o}dinger picture.
	
	We now turn to calculating the last two factors in \eqref{eqCorrRealization}. We consider the wavefunction $\ket{\psi_1(t)} \coloneqq \hat{U}_I(t,0) \ket{1}$, which obeys the equation of motion
	\begin{align}
	\iu \frac{\diff}{\diff t} \ket{\psi_1(t)} = \left(\hat{H}_0 + \hat{V}(t)\right) \ket{\psi_1(t)}.
	\label{eqSchrodinger}
	\end{align}
	The low frequency nature of the noise allows an adiabatic approximation to be applied, thanks to the finite energy gap of the system. Although the low energy subspace is degenerate, the symmetry properties of the Hamiltonian allow us to consider only the odd charge sector, so $\ket{\psi_1(t)}$ is always an odd-charge ground state of the instantaneous Hamiltonian:
	\begin{align}
	\ket{\psi_1(t)} = c_1(t) \ket{1_{\hat{V}(t)}},
	\end{align}
	where $\ket{1_{\hat{V}(t)}}$ is the unique odd-energy ground state of $(\hat{H}_0 + \hat{V}(t))$, which is normalized, so that $c_1(t)$ is a unit-modulus complex number. Although the overall phase of $\ket{1_{\hat{V}(t)}}$ is undetermined, physical quantities will be independent of any gauge transformations which change the phase of $\ket{1_{\hat{V}(t)}}$. We will find it useful later to adopt a specific gauge for $\ket{1_{\hat{V}(t)}}$, defined in the following way: We fix the phase of the unperturbed ground state $\ket{1}$, and define $\ket{1_{\hat{V}}}$ for a given perturbation $\hat{V}$ as the state that results from adiabatically evolving along the path $\hat{H}_0 + \lambda \hat{V}$, for $\lambda \in [0,1]$. Because the path has been specified, there is no longer a gauge ambiguity in the phase of $\ket{1_{\hat{V}}}$. The same can be defined for $\ket{0_{\hat{V}}}$. In this gauge, $\ket{1_{\hat{V}(t)}}$ is an implicit function of time only (there is no explicit $t$ dependence), and is differentiable in time (provided $\hat{V}(t)$ itself is differentiable).
	
	Inserting the adiabatic ansatz into \eqref{eqSchrodinger}, we find
	\begin{align}
	\iu \frac{\diff c_1(t)}{\diff t} = \epsilon(t) c_1(t) - \iu \Braket{1_{\hat{V}(t)} | \frac{\diff}{\diff t} 1_{\hat{V}(t)}}, \label{eqCoeffEom}
	\end{align}
	where $\epsilon(t)$ is the instantaneous energy of $\ket{1_{\hat{V}(t)}}$.
	
	The first term represents a dynamical phase, and can be eliminated by a parametrization
	\begin{align}
	c_1(t) = \tilde{c}_1(t) \exp\left(-\iu \int_0^t \diff t'\, \epsilon(t') \right),
	\label{eqDynamPhase}
	\end{align}
	such that $\tilde{c}_1(t)$ obeys the equation of motion \eqref{eqCoeffEom} without the first term on the right hand side.
	
	The last term of \eqref{eqCoeffEom} is identified as an Abelian Berry connection, which gives a geometric contribution to the phase of $\ket{\psi(t)}$. We will find it useful to write the connection in terms of the definition of the derivative
	\begin{align}
	\mathcal{A}_{11}(t) &\coloneqq \Braket{1_{\hat{V}(t)} | \frac{\diff}{\diff t} 1_{\hat{V}(t)}} \nonumber\\ &= \lim_{\delta t \rightarrow 0} \frac{\braket{1_{\hat{V}(t)}|1_{\hat{V}(t+\delta t)}} - 1}{\delta t},
	\label{eqConnection}
	\end{align}
	and $\mathcal{A}_{00}(t)$ is defined analogously. Normalization of $\ket{1_{\hat{V}(t)}}$ requires $\mathcal{A}_{11}^*(t) = -\mathcal{A}_{11}(t)$.
	The degeneracy of the instantaneous eigenstates ensures that the dynamical phases \eqref{eqDynamPhase} cancel. We then find that the correlation function for a particular noise realisation \eqref{eqCorrRealization} is given by the expression Eq.~\eqref{eqExpConnectionMain} in the main text. 
	
	
	We must now understand how the geometric phases transform under the action of symmetries, which will give us Eqs.~\eqref{eqConnSymUnitMain} and \eqref{eqConnSymAntiuMain}. Let us first consider the case of a unitary $\mathbb{Z}_2 \times \mathbb{Z}_2$ symmetry. From the form of the symmetry operators \eqref{eqSymOps}, and in the gauge chosen, we have $\ket{1_{\hat{V}(t)}} = \hat{\tau}^x \ket{0_{\hat{V}(t)}}$ for all time $t$ [where we understand the action of $\hat{\tau}^x$ to be extended to the whole Hilbert space in a way compatible with \eqref{eqSymOps}]. We have $\braket{1_{\hat{V}(t)}|1_{\hat{V}(t+\delta t)}} = \braket{0_{\hat{V}(t)}| \hat{\tau}^x \hat{\tau}^x | 0_{\hat{V}(t+\delta t)}}$, and $(\hat{\tau}^x)^2 = \mathbbm{1}$. From the definition \eqref{eqConnection}, this proves Eq.~\eqref{eqConnSymUnitMain}.
	
	In this case, the integral in the exponent in \eqref{eqExpConnectionMain} vanishes, and all the time-dependence is contained in the overlaps $\braket{0|0_{\hat{V}(t)}} \braket{1_{\hat{V}(t)}|1}$, which itself is equal to $|\braket{0|0_{\hat{V}(t)}}|^2$ by symmetry. In the gauge we have chosen, this overlap only depends on the value of $\hat{V}(t)$ at the final time, and is independent of $\hat{V}(s)$ for $s < t$. Therefore, if we expand this overlap as a perturbation series in $\hat{V}(t)$ and then average over noise realisations, the noise correlators will give factors of $\tilde{S}_\alpha(0)$, which is independent of $t$. We conclude that when the symmetries are unitary, the edge mode contribution to $\Gamma_B(t)$ takes the form
	\begin{align}
	\Gamma_B(t) &= |B_{01}|^2\left[1 - \mathcal{O}\left(\frac{V^2}{E_g^2}\right)\right] &(\mathbb{Z}_2 \times \mathbb{Z}_2)
	\end{align}
	where the correction is time-independent, and thus the sharp peak in the spectral function at $\omega = 0$ is preserved.
	
	We now turn to the antiunitary group $\mathbb{Z}_2 \times \mathbb{Z}_2^T$. In this case, the relationship between the two ground states takes the form $\ket{1_{\hat{V}(t)}} = \hat{\tau}^x \hat{\mathcal{K}} \ket{0_{\hat{V}(t)}}$, which differs from the unitary case by a complex conjugation (recall that $\hat{\mathcal{K}}$ is the complex conjugation operator acting to the right). We can ask what structure this symmetry imposes on the Berry connection. Applying this identity to the overlap factor in \eqref{eqConnection}, we find
	\begin{align}
	\braket{1_{\hat{V}(t)}|1_{\hat{V}(t+\delta t)}} &= \braket{0_{\hat{V}(t)}|\hat{\mathcal{K}}_{\leftarrow}\hat{\tau}^x \hat{\tau}^x \hat{\mathcal{K}} |0_{\hat{V}(t+\delta t)}^*} \nonumber\\
	&= \braket{0_{\hat{V}(t)}|0_{\hat{V}(t+\delta t)}}^* \nonumber\\
	\Rightarrow \mathcal{A}_{11}(t) &= -\mathcal{A}_{00}(t),
	\end{align}
	where $\hat{\mathcal{K}}_{\leftarrow}$ denotes the complex conjugation operator acting to the left. We have use the fact that $\mathcal{A}_{00}(t)$ is pure imaginary. This proves the relation \eqref{eqConnSymAntiuMain}.
	
	Since the geometric phases do not cancel when the symmetries are antiunitary, we must now calculate the correlator \eqref{eqExpConnectionMain}. To do this, we exploit a feature of the gauge we have chosen. A given noise realization $\hat{V}(s)$ ($0 \leq s \leq t$) can be represented as a trajectory in an $M$-dimensional parameter space, spanned by the coordinates $\vec{\eta}(s)$, which in turn determine $\hat{V}(s)$ via Eq.~\eqref{eqNoiseDecomp}. The state $\ket{\psi_1(t)}$ is the result of adiabatic deformation along this path, which starts at $\vec{\eta}(0) = \vec{0}$ by the assumption of adiabaticity \footnote{If the noise has a non-zero value at the initial time $t = 0$, one can redefine the gauge such that $\ket{1_{\hat{V}(t)}}$ is the result of adiabatic deformation under a straight line in parameter space starting at $\vec{\eta}(0)$}. On the other hand, by the definition of our gauge, the state $\ket{1_{\hat{V}(t)}}$ is the result of adiabatic deformation along a straight line in parameter space $\vec{\eta}(s) = (s/t) \vec{\eta}(t)$. The phase difference between the two states is given by the dynamical phase plus the difference of the geometric phases acquired along the two paths. Since both paths start at $\vec{\eta}(0) = \vec{0}$ and end at $\vec{\eta}(t)$, this phase difference can be expressed as a loop integral
	\begin{align}
	\theta^B_1(t) = -\iu \oint_{\mathcal{C}} \diff \vec{\eta} \cdot \vec{\mathcal{A}}_{11}(\vec{\eta})
	\label{eqBerry}
	\end{align}
	where the Berry connection in parameter space is given by $\vec{\mathcal{A}}_{00}(\vec{\eta}) = \braket{0_{\hat{V}(\eta)}|\vec{\nabla}_\eta 0_{\hat{V}(\eta)}}$, and the loop $\mathcal{C}$ is constructed as a  concatenation of the actual path taken by the noise $\vec{\eta}(s)$ with a straight line back to the origin. In terms of this gauge invariant Berry phase, the correlator for a particular noise realization is
	\begin{align}
	\Gamma_B[\hat{V}(s)](t) = |B_{01}|^2 e^{2\iu \theta^B_1(t)}.
	\label{eqGammaBerry}
	\end{align}
	The measured spectral function can be calculated by averaging the above over noise realizations, and then Fourier transforming.
	
	If the noise amplitude is small compared to the energy gap, we can estimate the noise average of $e^{2\iu \theta^B_1(t)}$ as a function of time. Using Stokes' theorem, the Berry phase \eqref{eqBerry} can be expressed as the flux of Berry curvature through a surface bounded by $\mathcal{C}$. For small noise amplitudes, the parameters $\vec{\eta}(s)$ explore only a small region around the origin, and in this region the Berry flux can be assumed to be a constant tensor $\Omega_{\alpha \beta}$. When there are two noise channels $M = 2$, the Berry phase is approximately given by the signed area swept out by $\eta(0 \leq s \leq t)$ (bounded by a straight line according to the choice of gauge) times the off diagonal value of the Berry curvature $\Omega_{xy}$. (The same results can be obtained for larger $M$ by projecting onto a two-dimensional surface orthogonal to the Berry flux.) This area can be calculated as
	\begin{align}
	A(t) = \int_0^t \diff s\, \eta_y \frac{\diff \eta_x }{\diff s}
	\end{align}
	where $\eta_{x,y}$ are the two components of $\vec{\eta}$. If the noise signals are Gaussian, then the average of $e^{2\iu \theta(t)}$ can be calculated through a cumulant expansion, which requires us to calculate the second moment of the swept area
	\begin{align}
	\overline{A(t)^2} &= \int_0^t \diff s \int_0^t \diff s' \overline{\eta_y(s) \eta_y(s') \frac{\diff \eta_x(s)}{\diff s} \frac{\diff \eta_x(s')}{\diff s'}} \nonumber\\
	&= \int_0^t \diff s \int_0^t \diff s' \tilde{C}_{xx}(s-s') \frac{\partial^2}{\partial s \partial s'} \tilde{C}_{yy}(s-s').
	\end{align}
	At times much longer than the noise correlation time, we can integrate over the sum and difference of $s$ and $s'$, discarding the corrections at the boundaries of the integral to give
	\begin{align}
	\overline{A(t)^2} &\approx \frac{1}{2} \int_0^{2t} \diff v \int_{-t}^t \diff u\, \tilde{C}_{xx}'(u) \tilde{C}_{yy}'(u).
	\end{align}
	We take the correlators $\tilde{C}_{\alpha \alpha}(t)$ to be on the order of $V^2$ times a dimensionless function of $(t/\tau_n)$, where $V$ is the typical strength of the noise and $\tau_n$ is the noise correlation time. This dimensionless function will be peaked with a width of order 1, such that the integral over $u$ in the above becomes independent of $t$. We therefore estimate the swept area to be
	\begin{align}
	\overline{A(t)^2} \sim \frac{tV^4}{\tau_n}.
	\end{align}
	Finally, we estimate the noise-averaged correlation function \eqref{eqGammaBerry}, which is give by $|B_{01}|^2 e^{-4\overline{A(t)^2}\Omega_{xy}}$
	\begin{align}
	\overline{\Gamma_B(t)} = |B_{01}|^2 e^{-\kappa tV^4/\tau_n E_g^4}
	\end{align}
	where $\kappa$ is a non-universal dimensionless constant that depends on the specific form of the spectral correlators. When this is Fourier transformed, we arrive at the spectral function Eq.~\eqref{eqSpecBroad}.

	\nocite{apsrev41Control}
	\bibliographystyle{apsrev4-1-prx}
	\bibliography{spt_ooe}
	
\end{document}